%% file: main.tex
\documentclass[prd, superscriptaddress, tightenlines, longbibliography, nofootinbib, eqsecnum, amsfonts, amsmath, amssymb, floatfix, notitlepage]{revtex4-2}
\pdfoutput=1
\usepackage[utf8]{inputenc}
\usepackage{amsmath}
\usepackage{mathtools}
\usepackage{amssymb}
\usepackage{graphics}
\usepackage{graphicx}
\usepackage{bm}
\usepackage[usenames,dvipsnames,svgnames,table]{xcolor}
\usepackage{appendix}
\usepackage{natbib}
\usepackage[breaklinks, colorlinks, urlcolor=urlcolor, linkcolor=linkcolor,citecolor=citecolor,pdfencoding=auto]{hyperref}
\usepackage{xspace}
\usepackage{float}
\usepackage{tikz}
\usepackage{multirow}
\usepackage{orcidlink}

\graphicspath{figures}

\DeclareMathAlphabet\mathbfcal{OMS}{cmsy}{b}{n}

\definecolor{linkcolor}{rgb}{0.6,0,0}
\definecolor{citecolor}{rgb}{0,0,0.75}
\definecolor{urlcolor}{rgb}{0.12,0.46,0.7}
\definecolor{Gray}{gray}{0.9}

\begin{document}
\include{aux_files/macros}
\include{texbase/macros}

\title{Constraints on $\taunl$ from \planck\ temperature and polarization}

\newcommand{\Sussex}{Department of Physics \& Astronomy, University of Sussex, Brighton BN1 9QH, United Kingdom}

\newcommand{\Geneve}{Universit\'e de Gen\`eve, D\'epartement de Physique Th\'eorique et CAP, 24 Quai Ansermet, CH-1211 Gen\`eve 4, Switzerland}

\author{$^{\text{\orcidlink{0000-0003-2060-8956}}}$Kareem Marzouk}
\affiliation{\Sussex}

\author{$^{\text{\orcidlink{0000-0001-5927-6667}}}$Antony Lewis}
\affiliation{\Sussex}

\author{$^{\text{\orcidlink{0000-0002-5751-1392}}}$Julien Carron}
\affiliation{\Geneve}
\affiliation{\Sussex}

\input{sections/abstract/abstract}

\date{\today}
\maketitle

\input{sections/introduction/introduction}
\input{sections/theory_and_methodology/theory_and_methodology}
\input{sections/data_and_sims/data_and_sims}
\input{sections/pipeline_validation/pipeline_validation}
\input{sections/results/results}

\input{sections/results/conclusions}
\input{sections/acknowledgements/acknowledgements}
\input{sections/appendix/appendix}
\bibliography{texbase/antony, texbase/cosmomc, aux_files/more_refs}

\allowdisplaybreaks

\end{document}

%% file: aux_files/macros.tex
\newcommand{\veck}{\mathbf{k}}

\newcommand{\WF}{{\rm WF}}

\newcommand{\bfdata}[1]{\mathbf{\hat{#1}}}
\newcommand{\bfspindata}[2]{{}_{#1} \mathbf{\hat{#2}}}

\newcommand{\rmdata}[1]{\mathrm{\hat{#1}}}
\newcommand{\rmspindata}[2]{{}_{#1} \mathrm{\hat{#2}}}

\newcommand{\bfcomp}[1]{\mathbf{#1}}
\newcommand{\bfspincomp}[2]{{}_{#1} \mathbf{#2}}

\newcommand{\rmcomp}[1]{\mathrm{#1}}
\newcommand{\rmspincomp}[2]{{}_{#1} \mathrm{#2}}

\newcommand{\transfer}{\mathbfcal{T}}

\newcommand{\field}[2]{
    \ensuremath{{#1}(#2)}
}

\newcommand{\vecl}[1]{
    \ensuremath{\textbf{l}_{#1}}
}

\newcommand{\todo}[1]{
    \textit{\color{red} #1}
}

\newcommand{\nzero}{
	\ifmmode {N_L^{(0)}}
	\else $N_L^{(0)}$
	\fi
}

\newcommand{\rdnzero}{\ensuremath{N_L^{{\rm RD},(0)}}}
\newcommand{\rdnzerobar}{\ensuremath{\bar{N}_L^{{\rm RD},(0)}}}
\newcommand{\rdnzerohat}{\ensuremath{\hat{N}_L^{{\rm RD},(0)}}}

\newcommand{\mcnzero}{\ensuremath{N_L^{{\rm MC},(0)}}}
\newcommand{\mcnzerobar}{\ensuremath{\bar{N}_L^{{\rm MC},(0)}}}
\newcommand{\mcnzerohat}{\ensuremath{\hat{N}_L^{{\rm MC},(0)}}}

\newcommand{\clzeta}{
	\ifmmode {C_L^{\zeta_*}}
	\else $C_L^{\zeta_*}$
	\fi
}

\newcommand{\deltalens}{\Delta \bar{C}_L^{(0)}}

\newcommand{\modf}{\ensuremath{f}}
\newcommand{\modfone}{\ensuremath{f^{X_1 Z_1}}}
\newcommand{\modftwo}{\ensuremath{f^{X_2 Z_2}}}

\newcommand{\clmodf}{\ensuremath{C_L^\modf}}
\newcommand{\clmodfoneone}{\ensuremath{C_L^{X_1 Z_1 \ X_1 Z_1}}}
\newcommand{\clmodftwotwo}{\ensuremath{C_L^{X_2 Z_2 \ X_2 Z_2}}}
\newcommand{\clmodfonetwo}{\ensuremath{C_L^{X_1 Z_1 \ X_2 Z_2}}}
\newcommand{\clmodftwoone}{\ensuremath{C_L^{X_2 Z_2 \ X_1 Z_1}}}

\newcommand{\clmodfhat}{\ensuremath{\hat{C}_L^\modf}}
\newcommand{\clmodfhatoneone}{\ensuremath{\hat{C}_L^{X_1 Z_1 \ X_1 Z_1}}}
\newcommand{\clmodfhattwotwo}{\ensuremath{\hat{C}_L^{X_2 Z_2 \ X_2 Z_2}}}
\newcommand{\clmodfhatonetwo}{\ensuremath{\hat{C}_L^{\mathcal{S}_1 X_1 Z_1, \ \mathcal{S}_2  X_2 Z_2}}}
\newcommand{\clmodfhattwoone}{\ensuremath{\hat{C}_L^{X_2 Z_2 \ X_1 Z_1}}}

\newcommand{\clmodfbar}{\ensuremath{\bar{C}_L^\modf}}
\newcommand{\clmodfbaroneone}{\ensuremath{\bar{C}_L^{X_1 Z_1 \ X_1 Z_1}}}
\newcommand{\clmodfbartwotwo}{\ensuremath{\bar{C}_L^{X_2 Z_2 \ X_2 Z_2}}}
\newcommand{\clmodfbaronetwo}{\ensuremath{\bar{C}_L^{\mathcal{S}_1 X_1 Z_1, \ \mathcal{S}_2  X_2 Z_2}}}
\newcommand{\clmodfbartwoone}{\ensuremath{\bar{C}_L^{X_2 Z_2 \ X_1 Z_1}}}

\newcommand{\flm}{\ensuremath{\modf_{LM}}}

\newcommand{\flmhat}{\ensuremath{{}_{\mathcal{S}}\hat{\modf}_{LM}^{X Z}}}
\newcommand{\flmhatone}{\ensuremath{{}_{\mathcal{S}_1}\hat{\modf}_{LM}^{X_1 Z_1}}}
\newcommand{\flmhattwo}{\ensuremath{{}_{\mathcal{S}_2}\hat{\modf}_{LM}^{X_2 Z_2}}}

\newcommand{\mf}{\ensuremath{{}_{\mathcal{S}}      f_{LM}^{X Z, \mathrm{MF}}}}
\newcommand{\mfone}{\ensuremath{{}_{\mathcal{S}_1} f_{LM}^{X_1 Z_1, \mathrm{MF}}}}
\newcommand{\mftwo}{\ensuremath{{}_{\mathcal{S}_2} f_{LM}^{X_2 Z_2, \mathrm{MF}}}}

\newcommand{\modfhat}{\ensuremath{\hat{\modf}}}

\newcommand{\npipe}{\texttt{NPIPE}\xspace}
\newcommand{\ffpten}{\texttt{FFP10}\xspace}
\newcommand{\dxtw}{\texttt{DX12}\xspace}
\newcommand{\prone}{\texttt{PR1}\xspace}
\newcommand{\prtwo}{\texttt{PR2}\xspace}
\newcommand{\prthree}{\texttt{PR3}\xspace}
\newcommand{\prfour}{\texttt{PR4}\xspace}
\newcommand{\splita}{\texttt{A}\xspace}
\newcommand{\splitb}{\texttt{B}\xspace}
\newcommand{\hmone}{\texttt{HM1}\xspace}
\newcommand{\hmtwo}{\texttt{HM2}\xspace}

\newcommand{\figref}[1]{Figure~\ref{#1}\xspace} 

%% file: texbase/macros.tex
\newcommand{\orcid}[1]{{\sc ORCiD:} \href{https://orcid.org/#1}{#1}}
\newcommand{\orcidacronym}[2]{\href{https://orcid.org/#2}{#1}}
\newcommand{\ALorcid}{\orcidacronym{AL}{0000-0001-5927-6667}}

\newcommand{\la}{\langle}
\newcommand{\ra}{\rangle}

\newcommand{\onesig}[1]{(68\%, \text{#1})}
\newcommand{\twosig}[1]{(95\%, \text{#1})}
\newcommand{\twoonesig}[3]{
\begin{equation}
\left.
 \begin{aligned}
#1 \\ #2
 \end{aligned}
\ \right\} \ \ \mbox{\text{#3}}
\end{equation}
}
\newcommand{\twotwosig}[3]{
\begin{equation}
\left.
 \begin{aligned}
#1 \\ #2
 \end{aligned}
\ \right\} \ \ \mbox{95\%, \text{#3}}
\end{equation}
}

\newcommand{\fsky}{f_{\rm sky}}

\providecommand{\planck}{\textit{Planck}}
\providecommand{\Planck}{\planck}
\newcommand{\camspec}{{\tt CamSpec}}
\newcommand{\plik}{{\tt Plik}}
\newcommand{\commander}{{\tt Commander}}
\newcommand{\mspec}{{\tt MSPEC}}
\newcommand{\smica}{{\tt SMICA}}

\newcommand{\MCNzero}{\textrm{MC-}N^{(0)}}
\newcommand{\RDNzero}{\textrm{RD-}N^{(0)}}
\newcommand{\RDNone}[0]{\textrm{RD-}N^{(1)}}
\newcommand{\xpRDNone}[0]{\textrm{xpRD-}N^{(1)}}
\newcommand{\MCNone}[0]{\textrm{MC-}N^{(1)}}

\newcommand{\CFHTLENS}{CFHTLenS}
\newcommand{\meanchisquare}{\overline{\chi^2}}

\newcommand{\mksym}[1]{\ifmmode {\rm #1}\else #1\fi}
\newcommand{\dataplus}{{+}}
\newcommand{\WP}{\mksym{WP}}
\newcommand{\highL}{\mksym{highL}}
\newcommand{\BAO}{\mksym{BAO}}
\newcommand{\lensing}{\mksym{lensing}}
\newcommand{\ext}{\mksym{ext}}
\newcommand{\planckonly}{\planck}
\newcommand{\TT}{\mksym{TT}}
\newcommand{\TTTEEE}{\mksym{TT,TE,EE}}
\newcommand{\planckTTonly}{\planck\ \TT}
\newcommand{\planckTTTEEEonly}{\planck\ \TTTEEE}
\newcommand{\lowTEB}{\mksym{lowP}}
\newcommand{\lowEB}{\mksym{lowP}}
\newcommand{\WMAPTEB}{\lowTEB\dataplus\mksym{WP}}
\newcommand{\lowTP}{\mksym{lowT,P}}
\newcommand{\planckTT}{\planckTTonly\dataplus\lowTEB}
\newcommand{\planckall}{\planckTTTEEEonly\dataplus\lowTEB}
\newcommand{\planckTTBAO}{\planckTT\dataplus\BAO}
\newcommand{\planckTTlensing}{\planckTT\dataplus\lensing}
\newcommand{\planckallBAO}{\planckall\dataplus\BAO}
\newcommand{\planckalllensing}{\planckall\dataplus\lensing}
\newcommand{\planckTTlensext}{\planckTT\dataplus\lensing\dataplus\ext}
\newcommand{\datalabel}[1]{#1}

\newcommand{\shortTT}{\TT\dataplus\lowTEB}
\newcommand{\shortall}{\TTTEEE\dataplus\lowTEB}

\newcommand{\HighL}{\highL}

\newcommand{\As}{A_{\rm s}}
\newcommand{\At}{A_{\rm t}}
\newcommand{\nt}{n_{\rm t}}
\newcommand{\ns}{n_{\rm s}}
\newcommand{\lcdm}{{$\rm{\Lambda CDM}$}}
\newcommand{\rpivot}{r_{0.05}}
\newcommand{\rzerotwo}{r_{0.002}}
\newcommand{\Alens}{A_{\rm L}}
\newcommand{\Aphiphi}{A_{\rm L}^{\phi\phi}}
\newcommand{\omegak}{\Omega_K}
\newcommand{\omegal}{\Omega_\Lambda}
\newcommand{\alphaiso}{\alpha}

\newcommand{\thetaMC}{\theta_{\rm MC}}
\newcommand{\nrun}{d \ns / d\ln k}
\newcommand{\nrunfrac}{\frac{d \ns}{d\ln k}}
\newcommand{\zre}{z_{\text{re}}}
\newcommand{\yhe}{Y_{\text{P}}}
\newcommand{\ypbbn}{Y_{\text{P}}^{\rm BBN}}
\newcommand{\zeq}{z_{\text{eq}}}
\newcommand{\nnu}{N_{\rm eff}}
\newcommand{\neff}{\nnu}
\newcommand{\mnu}{\sum m_\nu}
\newcommand{\sumnu}{\sum m_\nu}
\newcommand{\mnusterile}{m_{\nu,\, \mathrm{sterile}}^{\mathrm{eff}}}
\newcommand{\meffsterile}{\mnusterile}
\newcommand{\Tactive}{T_{\mathrm{a}}}
\newcommand{\Tsterile}{T_{\mathrm{s}}}
\newcommand{\msthermal}{m_{\rm sterile}^{\rm thermal}}
\newcommand{\msDW}{m_{\rm sterile}^{\rm DW}}

\newcommand{\zdrag}{z_{\rm drag}}
\newcommand{\rdrag}{r_{\rm drag}}
\newcommand{\zstar}{z_{\ast}}
\newcommand{\rstar}{r_{\ast}}
\newcommand{\rs}{r_{\rm s}}
\newcommand{\thetastar}{\theta_{\ast}}
\newcommand{\DAstar}{D_{\rm A}}
\newcommand{\keq}{k_{\rm eq}}
\newcommand{\Ailensbestfit}{A_i^{\mbox{\scriptsize{theory}}}}
\newcommand{\Alensbestfit}{A^{\mbox{\scriptsize{theory}}}}
\newcommand{\DVBAO}{D_{\rm V}}
\newcommand{\DA}{D_{\rm A}}

\newcommand{\fixme}[1]{{\color{Red}{ FIXME: #1}}}
\newcommand{\quietfixme}[1]{{\color{Red} #1}}
\newcommand{\notate}[1]{{\color{olive} NOTE: #1}}

\newcommand{\fnl}{\ensuremath{f_{\text{NL}}}}

\newcommand{\taunl}{\ensuremath{{\tau_{\text{NL}}}}}

\newcommand{\gnl}{\ensuremath{g_{\text{NL}}}}

\newcommand{\htaunl}{\ensuremath{{\hat{\tau}_{\rm NL}}}}

\newcommand{\fnlloc}{f^{\rm local}_{\rm NL}}
\newcommand{\fnleqi}{f^{\rm equil}_{\rm NL}}

\providecommand{\Planck}{\textit{Planck}}
\providecommand{\planck}{\Planck}
\newcommand{\covinvT}{\bar{\Theta}}
\newcommand{\Ltemp}{\tilde{T}}
\newcommand{\lensedC}{\tilde{C}}
\newcommand{\Lmin}{{L_{\rm min}}}

\providecommand{\lea}{\la}
\providecommand{\gea}{\ga}

\providecommand{\alt}{\lea}
\providecommand{\agt}{\gea}
\providecommand{\text}[1]{\rm{#1}}

\newcommand{\Msun}{M_\odot}
\newcommand{\tot}{{\text{tot}}}
\newcommand{\Mpc}{\text{Mpc}}
\newcommand{\half}{{\textstyle \frac{1}{2}}}
\newcommand{\third}{{\textstyle \frac{1}{3}}}
\newcommand{\numfrac}[2]{{\textstyle \frac{#1}{#2}}}
\renewcommand{\d}{\text{d}}
\newcommand{\grad}{\nabla}
\newcommand{\km}{{\rm{\,km\,}}}

\newcommand{\Hunit}{\text{km}\,\text{s}^{-1}\,\Mpc^{-1}}
\newcommand{\Gyr}{{\rm Gyr}}
\providecommand{\muK}{\mu{\rm K}}
\providecommand{\arcmin}{{\rm arcmin}}
\newcommand{\muKarcmin}{\,\muK\,\arcmin}
\newcommand{\muKsq}{(\mu\rm{K})^2}
\newcommand{\lmax}{l_{\text{max}}}
\newcommand{\lmin}{l_{\text{min}}}

\newcommand{\mpl}{m_{\text{Pl}}}
\newcommand{\eV}{\,\text{eV}}
\newcommand{\MeV}{\,\text{MeV}}
\newcommand{\GeV}{\,\text{GeV}}

\providecommand{\Omk}{\Omega_K}
\providecommand{\Oml}{\Omega_{\Lambda}}
\providecommand{\Omtot}{\Omega_{\mathrm{tot}}}
\providecommand{\Omb}{\Omega_{\mathrm{b}}}
\providecommand{\Omc}{\Omega_{\mathrm{c}}}
\providecommand{\Omm}{\Omega_{\mathrm{m}}}
\providecommand{\omb}{\omega_{\mathrm{b}}}
\providecommand{\omc}{\omega_{\mathrm{c}}}
\providecommand{\omm}{\omega_{\mathrm{m}}}
\providecommand{\Omdm}{\Omega_{\mathrm{DM}}}
\providecommand{\Omnu}{\Omega_{\nu}}

\providecommand{\CAMB}{{\tt camb}}
\providecommand{\GetDist}{{\tt GetDist}}

\providecommand{\COSMOMC}{{\tt CosmoMC}}
\providecommand{\CMBFAST}{{\tt cmbfast}}
\providecommand{\COSMICS}{{\tt Cosmics}}
\providecommand{\CLASS}{{\tt class}}
\providecommand{\CMBEASY}{{\tt cmbeasy}}
\providecommand{\LCDM}{{$\rm{\Lambda CDM}$}}
\providecommand{\COSMOREC}{{\tt CosmoRec}}
\providecommand{\HYREC}{{\tt HyRec}}
\providecommand{\RECFAST}{{\tt recfast}}
\providecommand{\HALOFIT}{{\tt halofit}}
\providecommand{\AlterBBN}{{\tt AlterBBN}}
\providecommand{\MontePython}{{\tt Monte Python}}
\providecommand{\HEALpix}{{\tt HEALpix}}

\newcommand{\begm}{\begin{pmatrix}}
\newcommand{\enm}{\end{pmatrix}}
\newcommand{\threej}[6]{{\begm #1 & #2 & #3 \\ #4 & #5 & #6 \enm}}
\newcommand{\threejz}[3]{{\begm #1 & #2 & #3 \\ 0 & 0 & 0 \enm}}

\newcommand\ba{\begin{eqnarray}}
\newcommand\ea{\end{eqnarray}}
\newcommand\bea{\begin{eqnarray}}
\newcommand\eea{\end{eqnarray}}

\newcommand\be{\begin{equation}}
\newcommand\ee{\end{equation}}

\newcommand{\valpha}{{\boldsymbol{\alpha}}}
\newcommand{\vgrad}{{\boldsymbol{\nabla}}}
\newcommand{\vpsi}{\mathbf{\psi}}
\newcommand{\vTheta}{\mathbf{\Theta}}
\newcommand{\vtheta}{\boldsymbol{\theta}}
\newcommand{\vdelta}{\boldsymbol{\delta}}
\newcommand{\vell}{{\boldsymbol{\ell}}}


\providecommand{\var}{\text{var}}
\providecommand{\cov}{\text{cov}}

\providecommand{\Tr}{\text{Tr}}

\newcommand{\ud}{{\rm d}}

\newcommand{\mA}{\bm{A}}
\newcommand{\mB}{\bm{B}}
\newcommand{\mC}{\bm{C}}
\newcommand{\mD}{\bm{D}}
\newcommand{\mE}{\bm{E}}
\newcommand{\mF}{\bm{F}}
\newcommand{\mG}{\bm{G}}
\newcommand{\mH}{\bm{H}}
\newcommand{\mI}{\bm{I}}
\newcommand{\mJ}{\bm{J}}
\newcommand{\mK}{\bm{K}}
\newcommand{\mL}{\bm{L}}
\newcommand{\mM}{\bm{M}}
\newcommand{\mN}{\bm{N}}
\newcommand{\mO}{\bm{O}}
\newcommand{\mP}{\bm{P}}
\newcommand{\mQ}{\bm{Q}}
\newcommand{\mR}{\bm{R}}
\newcommand{\mS}{\bm{S}}
\newcommand{\mT}{\bm{T}}
\newcommand{\mU}{\bm{U}}
\newcommand{\mV}{\bm{V}}
\newcommand{\mW}{\bm{W}}
\newcommand{\mX}{\bm{X}}
\newcommand{\mY}{\bm{Y}}
\newcommand{\mZ}{\bm{Z}}

\newcommand{\mLambda}{\bm{\Lambda}}
\newcommand{\mzero}{\bm{0}}


\newcommand{\boldvec}[1]{{\mbox{\boldmath{$#1$}}}}

\newcommand{\vA}{\boldvec{A}}
\newcommand{\vB}{\boldvec{B}}
\newcommand{\vC}{\boldvec{C}}
\newcommand{\vD}{\boldvec{D}}
\newcommand{\vE}{\boldvec{E}}
\newcommand{\vF}{\boldvec{F}}
\newcommand{\vG}{\boldvec{G}}
\newcommand{\vH}{\boldvec{H}}
\newcommand{\vI}{\boldvec{I}}
\newcommand{\vJ}{\boldvec{J}}
\newcommand{\vK}{\boldvec{K}}
\newcommand{\vL}{\boldvec{L}}
\newcommand{\vM}{\boldvec{M}}
\newcommand{\vN}{\boldvec{N}}
\newcommand{\vO}{\boldvec{O}}
\newcommand{\vP}{\boldvec{P}}
\newcommand{\vQ}{\boldvec{Q}}
\newcommand{\vR}{\boldvec{R}}
\newcommand{\vS}{\boldvec{S}}
\newcommand{\vT}{\boldvec{T}}
\newcommand{\vU}{\boldvec{U}}
\newcommand{\vV}{\boldvec{V}}
\newcommand{\vW}{\boldvec{W}}
\newcommand{\vX}{\boldvec{X}}
\newcommand{\vY}{\boldvec{Y}}
\newcommand{\vZ}{\boldvec{Z}}

\newcommand{\va}{\boldvec{a}}
\newcommand{\vb}{\boldvec{b}}
\newcommand{\vc}{\boldvec{c}}
\newcommand{\vd}{\boldvec{d}}
\newcommand{\ve}{\boldvec{e}}
\newcommand{\vf}{\boldvec{f}}
\newcommand{\vg}{\boldvec{g}}
\newcommand{\vh}{\boldvec{h}}
\newcommand{\vi}{\boldvec{i}}
\newcommand{\vj}{\boldvec{j}}
\newcommand{\vk}{\boldvec{k}}
\newcommand{\vl}{\boldvec{l}}
\newcommand{\vm}{\boldvec{m}}
\newcommand{\vn}{\boldvec{n}}
\newcommand{\vo}{\boldvec{o}}
\newcommand{\vp}{\boldvec{p}}
\newcommand{\vq}{\boldvec{q}}
\providecommand{\vr}{\boldvec{r}}
\newcommand{\vs}{\boldvec{s}}
\newcommand{\vt}{\boldvec{t}}
\newcommand{\vu}{\boldvec{u}}
\newcommand{\vv}{\boldvec{v}}
\newcommand{\vw}{\boldvec{w}}
\newcommand{\vx}{\boldvec{x}}
\newcommand{\vy}{\boldvec{y}}
\newcommand{\vz}{\boldvec{z}}

\newcommand{\cla}{\mathcal{A}}
\newcommand{\clb}{\mathcal{B}}
\newcommand{\clc}{\mathcal{C}}
\newcommand{\cld}{\mathcal{D}}
\newcommand{\cle}{\mathcal{E}}
\newcommand{\clf}{\mathcal{F}}
\newcommand{\clg}{\mathcal{G}}
\newcommand{\clh}{\mathcal{H}}
\newcommand{\cli}{\mathcal{I}}
\newcommand{\clj}{\mathcal{J}}
\newcommand{\clk}{\mathcal{K}}
\newcommand{\cll}{\mathcal{L}}
\newcommand{\clm}{\mathcal{M}}
\newcommand{\cln}{\mathcal{N}}
\newcommand{\clo}{\mathcal{O}}
\newcommand{\clp}{\mathcal{P}}
\newcommand{\clq}{\mathcal{Q}}
\newcommand{\clr}{\mathcal{R}}
\newcommand{\cls}{\mathcal{S}}
\newcommand{\clt}{\mathcal{T}}
\newcommand{\clu}{\mathcal{U}}
\newcommand{\clv}{\mathcal{V}}
\newcommand{\clw}{\mathcal{W}}
\newcommand{\clx}{\mathcal{X}}
\newcommand{\cly}{\mathcal{Y}}
\newcommand{\clz}{\mathcal{Z}}

\newcommand{\vnhat}{\hat{\vn}}
\newcommand{\vrhat}{\hat{\vr}}
\newcommand{\vkhat}{\hat{\vk}}

\providecommand{\ltsima}{\lea}
\providecommand{\gtsima}{\gea}
\providecommand{\simlt}{\lea}
\providecommand{\simgt}{\gea}

\newcommand{\elec}{{\rm e}}
\newcommand{\neutron}{{\rm n}}

%% file: sections/abstract/abstract.tex
\begin{abstract}
We update constraints on the amplitude of the primordial trispectrum, using the final \planck\ mission temperature and polarization data. In the squeezed limit, a cosmological local trispectrum would be observed as a spatial modulation of small-scale power on the CMB sky. We reconstruct this signal as a source of statistical anisotropy via quadratic estimator techniques. We systematically demonstrate how the estimated power spectrum of a reconstructed modulation field can be translated into a constraint on \taunl\ via likelihood methods, demonstrating the procedures effectiveness by inferring known \taunl\ signal(s) from simulations. Our baseline results constrain $\taunl < 1700$ at the 95\% confidence level, providing the most stringent constraints to date.
\end{abstract} 

%% file: sections/introduction/introduction.tex
\section{Introduction}
A wide range of observations appear to be broadly consistent with a Universe that started the hot Big Bang phase with purely Gaussian adiabatic super-horizon scalar perturbations in an otherwise Friedmann-Robertson-Walker background. Gaussian adiabatic perturbations have only one locally varying degree of freedom (`single clock'), which is not locally observable on super-horizon scales since it corresponds to a local re-definition of the time coordinate or scale factor~\cite{Maldacena:2002vr,Creminelli:2004pv,Senatore:2012nq}.
Gaussian adiabatic models are simple and highly predictive, for example, the small-scale CMB perturbations in each independent Hubble patch at recombination are expected to be statistically identical. This would be consistent with inflation, where perturbation generation is dominated by the effect of a single effective scalar field. Any deviation from this simple prediction would be powerful evidence for a more complex multi-field inflationary evolution, or other new early universe physics involving more than one local degree of freedom. In this paper we constrain large-scale modulation of small-scale CMB fluctuations by reconstructing the modulation field using the latest \Planck\ data.

In the language of primordial non-Gaussianity, a large-scale modulation of small-scale power corresponds to `squeezed'-shape non-Gaussianity (see Ref.~\cite{Lewis:2011au,Byrnes:2010em,Bartolo:2004if} for reviews). We model the initial non-Gaussian curvature perturbation $\zeta$ as a locally-modulated version of an underlying Gaussian field $\zeta_0$, so that in real space
\begin{equation}
  \zeta(\vx) \approx \zeta_0(\vx)(1+\phi(\vx)),
\end{equation}
where $\phi$ is some modulating field, and we assume $|\phi(\vx)|\ll 1$. If $\phi$ is correlated to $\zeta_0$, there will be a non-zero local (squeezed) curvature bispectrum, as commonly parameterized by $\fnl$, corresponding in the CMB to the amplitude of the small-scale CMB power spectrum varying in a way that's correlated to the large-scale temperature.
However, whether or not $\phi$ is correlated to $\zeta_0$, there will always be a non-zero squeezed trispectrum, corresponding to the small-scale $\zeta$ power varying spatially with $\phi$. In the inflationary context this is usually parameterized by $\taunl$, where
$\taunl \geq \left(\frac{5}{6} \fnl\right)^2$ depending on the degree of correlation of $\phi$ and $\zeta_0$~\cite{Suyama:2007bg}. The local bispectrum constraint $\fnl=-0.9\pm 5.1$ from \Planck~\cite{Planck:2019kim} means that any $\taunl$ from a nearly scale-invariant $\phi\propto \zeta_0$ must be very small, however there remains the possibility of an uncorrelated modulation giving rise to a more substantial $\taunl$.

The best current empirical constraint comes from the \Planck\ 2013 nominal-mission temperature data, $\taunl < 2800\, (95\% {\rm CL})$~\cite{Planck2013-NG}, where $\taunl \sim 500$ would correspond to a modulation at the $|\phi|\sim 10^{-3}$ level\footnote{We quote constraints assuming no other non-Gaussian signals apart from CMB lensing (and other effects modelled in the simulations) are present. Ref.~\cite{Feng:2015pva} find a weaker joint constraint with $\gnl$, but also used a more suboptimal estimator.}. This result was a conservative upper limit, accounting for evidence of frequency-dependent foregrounds or systematic modulation in power (particularly in the octopole). With full-mission \Planck\ data~\cite{Planck:2018nkj} the temperature noise is substantially reduced and the CMB polarization data give additional constraining power; combined with better cleaning of foregrounds and modelling of systematics, it should therefore be possible to substantially improve on the first \Planck\ constraint. The updated \npipe analysis pipeline~\cite{Planck:2020olo} is also now available using slightly more data, lowering the noise further, as well as offering the prospect of a better control of systematics and more reliable simulations.

In practice, only large-scale modes of the modulation field can be constrained well, since small-scale variations in power are impossible to distinguish from random fluctuations in local realizations of the modes~\cite{Kogo:2006kh,Pearson:2012ba}. We focus on modulation modes corresponding to multipoles $L \alt 100$, where at recombination the modulation field remains largely super-horizon and approximately constant through the last-scattering surface. In this case, a primordial curvature modulation translates directly into a large-scale modulation of the small-scale CMB~\cite{Pearson:2012ba,Planck2013-NG}
\begin{equation}\label{Xmodel}
 X(\vnhat) \approx X_g(\vnhat)[1 + \phi(\vnhat, r_*)] \equiv X_g(\vnhat)[1 + \modf(\vnhat)],
\end{equation}
where $X$ is the small-scale CMB temperature perturbation or polarization field, $X_g$ is the corresponding standard unmodulated linear Gaussian field, and $r_*$ is the radial distance to last scattering. The two-dimensional modulation field $\modf(\vnhat)\equiv \phi(\vnhat, r_*)$ parameterizes the modulation directly in terms of the CMB observables, and can be reconstructed from the data using quadratic estimator techniques~\cite{Hanson:2009gu}. For measuring $\taunl$, where most of the signal is in modulation modes $L\ll 100$, the super-horizon approximation is accurate at the percent level, allowing a greatly simplified analysis~\cite{Pearson:2012ba} that is nearly optimal. With no additional computational cost we can also constrain $\modf({\vnhat})$ more generally, without assuming $\taunl$ form, allowing for more general scale dependence of the modulation field. The analysis is then equivalent to looking for a power modulation, as possibly suggested by hints for statistical anisotropy~\cite{Hansen:2004vq,Hanson:2009gu,Contreras:2017qkc,Akrami:2019bkn}. We shall however assume that all the small-scale modes are modulated as implied by the model of Eq.~\eqref{Xmodel}, rather than allowing for $X_g$ to be modulated in a different way depending on the CMB harmonic scale. This allows us to place good constraints using the full power of high-resolution \Planck\ data, and avoids issues regarding possible a posteriori choices for more general parameterizations (for a discussion of scale-dependent models, see e.g., Ref.~\cite{Byrnes:2010ft,Adhikari:2018osh}).

The signal modulation of Eq.~\eqref{Xmodel} is of exactly the same structure as having an optical depth to last scattering ($\Delta\tau$) that varies with line of sight, with
\begin{equation}
 X(\vnhat) = e^{-\Delta\tau(\vnhat)} X_g(\vnhat) \approx  X_g(\vnhat) [1-\Delta\tau(\vnhat)].
\end{equation}
The modulation reconstruction can therefore also be used to constrain patchy reionization~\cite{Dvorkin:2008tf,Dvorkin:2009ah,Gluscevic:2012qv,Namikawa:2017uke,Namikawa:2021zhh,Roy:2022muv}. The main difference is that the standard $\taunl$ model imposes a specific scale-dependence to the modulation power spectrum, where the signal of interest is concentrated on very large scales. Optical-depth perturbations by contrast have signal out to small scales, since the modulation more closely follows the density spectrum than the curvature/potential spectrum. Interpretation of the patchy reionization constraint is also more complicated as it depends on the details of the reionization model. There is also a significant correlation with CMB lensing~\cite{Feng:2018eal,Fidler:2019mny}, and polarization can also be generated by scattering at reionization. In this work we focus on constraining $\taunl$, but our full modulation reconstruction power spectrum (which is dominated by the temperature modulations) demonstrates a non-detection which could also be translated into a constraint on specific reionization models. For Planck-level sensitivity, the patchy reionization signal is not expected to significantly contaminate $\taunl$ at the level that could be detected.

%% file: sections/theory_and_methodology/theory_and_methodology.tex
\section{Theory and Methodology}
\input{sections/theory_and_methodology/subsection_squeezed_trispectrum}
\input{sections/theory_and_methodology/subsection_modulation_estimator}
\input{sections/theory_and_methodology/subsection_power_spectrum_estimator}
\input{sections/theory_and_methodology/subsection_likelihood} 

%% file: sections/theory_and_methodology/subsection_squeezed_trispectrum.tex
\subsection{A squeezed $\taunl$ trispectrum}\label{section:squeezed_taunl}

We construct the general trispectrum from four wavevectors $\vecl{1}, \vecl{2}, \vecl{3}, \vecl{4}$ that define a closed quadrilateral in harmonic space. We will consider $X=X(\vecl{})$ to be a random field, which may correspond to, for example, the temperature fluctuations of the CMB at last scattering. The four-point function $\langle X(\vecl{1}) X(\vecl{2}) X(\vecl{3}) X(\vecl{4})\rangle$ will receive two contributions, which are referred to as ``connected'' and ``disconnected'' pieces. Regardless of the statistical properties of the field $X$, the disconnected component can be factorized into the products of two-point function $\langle X(\vecl{i}) X(\vecl{j})\rangle$, i.e., power spectra, which are typically non-zero even if $X$ is a Gaussian random field. By comparison, the connected part of the four-point function is exactly zero when $X$ is Gaussian. We can therefore probe the non-Gaussianity of the field $X$ by analysing the connected trispectrum.

In this work, we consider a diagonally squeezed form of the trispectrum. To understand this, let us first consider a triangle in harmonic space. If we take this triangle, and squeeze one of its sides such that $|\mathbf{L}|\ll|\vecl{1}|, |\vecl{2}|$ and $|\mathbf{L}| \approx |\vecl{1} + \vecl{2}|$, then we necessarily have one large-scale mode and two small-scale modes. If the bispectrum positive, in regions of space with a positive large-scale mode, the small-scale power over those regions must be enhanced; correspondingly, it must be suppressed if the bispectrum is negative; and vice versa for a negative large-scale mode. The small-scale power therefore appears to be modulated by the large-scale mode, with a correlation between the small-scale power and the large-scale mode.

\begin{figure}
	\includegraphics[width=0.65\textwidth]{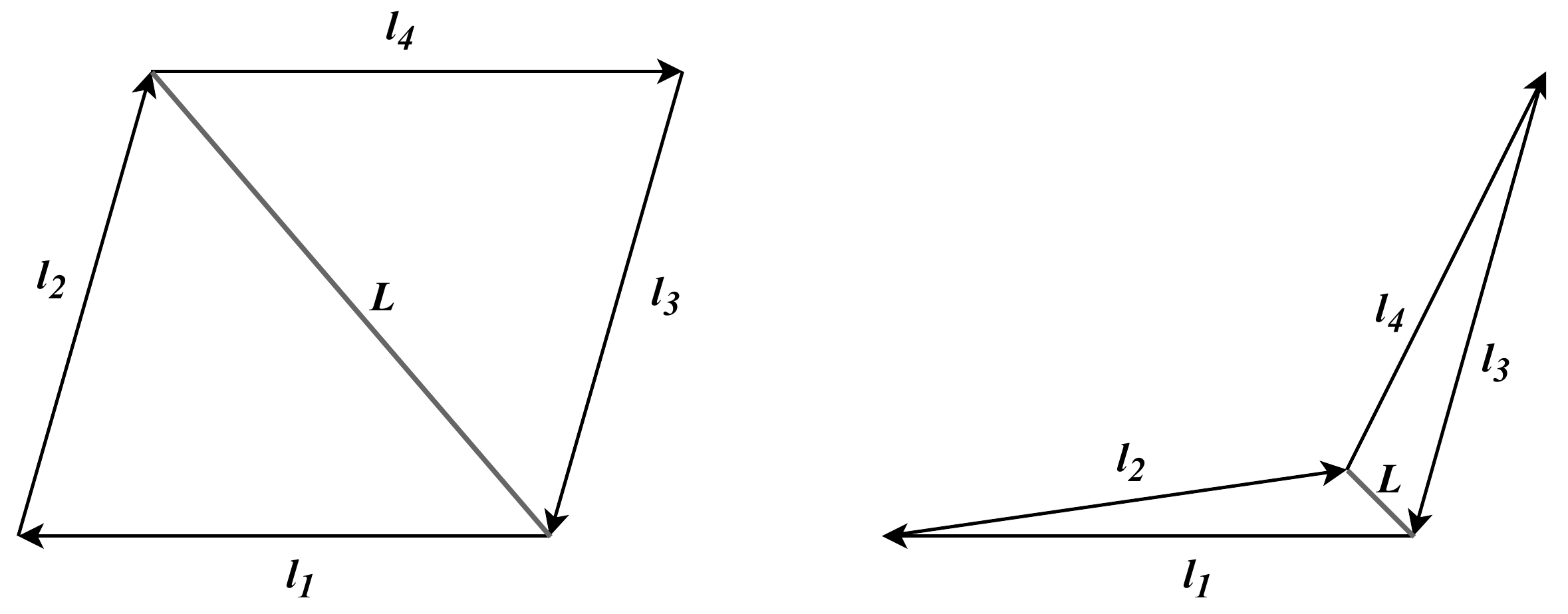}
	\caption{\textit{Left:} Example trispectrum. Wavevectors form a closed quadrilateral in harmonic space. The quadrilateral can be described equivalently by two triangles related by a common wavevector $L$. \textit{Right:} Schematic of a squeezed trispectrum, corresponding to the $\taunl$ signal peak; $L\ll l_1, l_2, l_3, l_4$ and $L\approx |\mathbf{l}_1 + \mathbf{l}_2| \approx |\mathbf{l}_3 + \mathbf{l}_4|$.
}
\label{fig:sq_quad}
\end{figure}

By extension to this triangular picture, we could append two further modes ($\vecl{3},\vecl{4}$) to the squeezed side of the triangle to form a closed quadrilateral. So long as these additional modes are much longer than the common diagonal length, $|\mathbf{L}| \ll |\vecl{3}|, |\vecl{4}|$, then we will form a diagonal-squeezed trispectrum. We display this geometry in Figure~\ref{fig:sq_quad}. In this case, the field with large-scale mode $\vL$ is not one of the observed modes defining the measured trispectrum, and does not have to be directly observable itself. If there is a correlation between the large-scale mode and the curvature perturbation, there will potentially be an observable bispectrum. However, the large-scale mode may be uncorrelated, or weakly correlated, in which case there may be a substantial trispectrum without a corresponding observable bispectrum.

If the large-scale mode is associated with a scalar field, as expected for example in simple multi-field inflation models, the squeezed-diagonal trispectrum is independent of the relative orientation of the small and large-scale modes. The small-scale power is therefore expected to remain locally isotropic, but to have a spatial modulation determined by the large-scale mode. Since the large-scale mode may not be observable, the connected trispectrum is what quantifies this observable modulation in small-scale power.

This squeezed diagonal trispectrum is what is known as the $\taunl$-shape. Its definition is usually accompanied with another shape constructed via squeezing one outer length of the quadrilateral such that $|\vecl{1}|\ll, |\vecl{2}|, |\vecl{3}|, |\vecl{4}|$, rather than across its diagonal. This ``$\gnl$-form'' trispectrum produces a large-scale modulation of a small-scale bispectrum~\cite{Lewis:2011au}. Together, we conventionally write the connected trispectrum as the sum
\begin{equation}\label{eq:connected_trispectrum}
    \langle X(\vecl{1}) X(\vecl{2}) X(\vecl{3}) X(\vecl{4})\rangle_c
    	=
     \langle X(\vecl{1}) X(\vecl{2}) X(\vecl{3}) X(\vecl{4})\rangle^{\gnl}_c
     +
     \langle X(\vecl{1}) X(\vecl{2}) X(\vecl{3}) X(\vecl{4})\rangle^{\taunl}_c.
\end{equation}
The $\taunl$ and $\gnl$ trispectra can be produced by local quadratic and cubic corrections to an otherwise Gaussian initial perturbation field. It is corrections of this type that are computed in the $\delta N$ formalism for inflationary perturbations~\cite{Byrnes:2006vq, Byrnes:2010ft, Byrnes:2010em, Smidt:2010ra}, and these are often used as key predictors for models of inflation described by multiple fields. The fact that the peak signals for $\taunl$ and $\gnl$ are attained in different momentum configurations allow us to form constraints almost independently of one another. In the remainder of this paper, we will only consider the $\taunl$ trispectrum, and refer the reader to the 2018 constraints on $\gnl$ by the \Planck\ collaboration~\cite{Planck:2019kim}.
It is usually also assumed that the large-scale modes are approximately scale invariant (have a power spectrum proportional to $P_{\zeta_0}$, as would be the case for local $\zeta_0^2$ and $\zeta_0^3$ corrections), which then determines the overall wavenumber-dependence of the signals.

The explicit form of the $\taunl$ trispectrum in Eq.~\eqref{eq:connected_trispectrum} connects the primordial fluctuations to those on the last scattering surface. We can write the trispectrum in its reduced form (i.e. with redundant symmetries eliminated) as~\cite{Okamoto:2002ik}
\begin{equation}\label{eq:transfer_taunl}
	p^{l_1 l_2}_{l_3 l_4}(L)
	=
	\taunl
	\int \d r_1 \d r_2 r_1^2 r_2^2 F_L(r_1, r_2) \Xi(r_1, l_1, l_2) \Xi(r_2, l_3, l_4),
\end{equation}
where $F_L(r_1, r_2) = 4\pi \int \d \ln k \mathcal{P}_\zeta (k)j_L(kr_1) j_L(kr_2)$ and $\Xi$ contains the radiative transfer functions and further spherical Bessel functions $j_L$ required for projecting onto the CMB surface. Exact results can be obtained numerically via Boltzmann codes, but there are analytic simplifications. In particular at recombination the kernel $F(r_1=r_*, r_2=r_*)=C_L^{\zeta_*}$ and in the squeezed limit the (approximate) constancy of the large-scale mode $L$ relative to the short scale fluctuations means that $\int \d r r^2  \Xi(r, l_i, l_j)\approx (C_{l_i} + C_{l_j})$~\cite{Pearson:2012ba}. The squeezed $\taunl$ trispectrum at recombination is therefore approximated as
\begin{equation}
	p^{l_1 l_2}_{l_3 l_4}(L)
	\approx
	\taunl C_L^{\zeta_*} (C_{l_1} + C_{l_2})(C_{l_3} + C_{l_4}).
\end{equation}
This is exactly the form of squeezed trispectrum expected from the modulated Gaussian field model of Eq.~\eqref{Xmodel}, where the modulation field is proportional to the curvature perturbation at recombination,  $f\propto \zeta_*$.

%% file: sections/theory_and_methodology/subsection_modulation_estimator.tex
\subsection{The modulation field estimator}\label{sec:modulation_estimator}
We denote the observed CMB temperature and polarization fields with the block vector
\begin{equation}\label{eq:fields}
	\bfdata{X} =
	(\bfspindata{0}{X}, \
	 \bfspindata{+2}{X}, \
	 \bfspindata{-2}{X}
	 )^T =
	(
	\bfspindata{}{T}, \
	\bfspindata{+2}{P}, \
	\bfspindata{-2}{P})^\mathrm{T},
\end{equation}
where we have adopted the use of ``spin indices'' to indicate the spin-0 temperature field $\bfspindata{}{T}$ and spin-2 polarization fields $\bfspindata{\pm 2}{P} = \bfdata{Q} \pm i \bfdata{U}$ in order to write the fields in a unified notation. CMB fields that appear in bold typeface refer to the vector of pixel-data, and if non-bold refer to their scalar value when evaluated at a specific location on the sky,
e.g. $\rmspindata{}{T}(\vnhat) \in \bfspincomp{}{T}$.
$\bfdata{Q}$ and $\bfdata{U}$ are the Stokes polarization maps which are obtained at the detector level. %
Each data component in Eq.~\eqref{eq:fields} is modelled as the linear combination of the cosmological signal and noise $\bfdata{X} = \bfcomp{X} + \bfcomp{N}$, each of which have the same block vector structure: $\bfcomp{X} = (\bfspincomp{0}{X}, \ \bfspincomp{+2}{X}, \ \bfspincomp{-2}{X})^\mathrm{T}$,  $\bfcomp{N} = (\bfspincomp{0}{N}, \ \bfspincomp{+2}{N}, \ \bfspincomp{-2}{N})^\mathrm{T}$. %
A large-scale modulation that is approximately constant through the last-scattering light cone will alter each component of the signal vector in the same way, which we define as a linear real-space operation
$\rmspincomp{s}{X}(\vnhat) = [1+f(\vnhat)]\rmspincomp{s}{X}_0(\vnhat)$ (where the modulation acting on the otherwise unmodulated signal
$\bfcomp{X}_0=(\bfspincomp{0}{X}_0, \ \bfspincomp{+2}{X}_0, \ \bfspincomp{-2}{X}_0)^\mathrm{T}$).
We then model the observed data as
\begin{equation}
	\bfdata{X} =  \mathbfcal{B}\mathbf{X} + \mathbf{N},
\end{equation}
where $\mathbfcal{B}$ denotes the operation of the instrumental beam (a convolution in real space) and pixel window function.

Assuming that the unmodulated signal and noise components are reliably approximated as Gaussian random fields, we can use maximum likelihood estimates (MLE) to build estimates of $f$. We follow the methods developed by Hirata and Seljak~\cite{Hirata:2002jy} (for CMB lensing), and later by Hanson and Lewis~\cite{Hanson:2009gu} (for generalized weak statistical anisotropy), defining the (negative) log-likelihood function
\begin{align}\label{eq:negloglike}
	\begin{split}
		\mathcal{L}[\mathbf{f}]
		&\equiv  -\ln P(\bfdata{X} | \mathbf{f})\\
		&=
		\frac{1}{2}
		\bfdata{X}^\dagger
		\left(\mathbf{C^{\hat{X}\hat{X}}}\right)^{-1}
		\bfdata{X}
		+
		\frac{1}{2}
		\ln
		\mathrm{det} \mathbf{C^{\hat{X}\hat{X}}} + \text{const}.
	\end{split}
\end{align}
The conditional probability $P(\bfdata{X} | \mathbf{f})$ appearing in the log-likelihood is the standard multivariate Gaussian density with data covariance $\mathbf{C^{\hat{X}\hat{X}}}$, which is not isotropic as it depends on the realization of the statistical anisotropy field $\mathbf{f}$. Note that we have represented the modulation $f$ as the vector $\mathbf{f}$ (here containing only one item) for notational compatibility with the formalism.

We can find $\mathbf{f}$ corresponding to the maximum-likelihood point by identifying the minimum of Eq.~\eqref{eq:negloglike}. This can be computed directly by finding where the functional derivative of the likelihood function is zero, $\delta \mathcal{L}/\delta \mathbf{f^\dagger} = 0$. As explicitly given in~\cite{Hanson:2009gu}, the derivative may be conveniently expressed in terms of the combination
\begin{equation}\label{eq:log_like_elements}
	\mathbfcal{H} =
	\frac{1}{2}
	\left[
	\mathbfcal{B}^\dagger
	\left( \mathbf{C^{\hat{X}\hat{X}}}\right)^{-1}
	\bfdata{X}
	\right]^\dagger
	\frac{\mathbf{\delta C^{\hat{X}\hat{X}}}}{\delta \mathbf{f^\dagger}}
	\left[
	\mathbfcal{B}^\dagger
	\left(\mathbf{C^{\hat{X}\hat{X}}}\right)^{-1}
	\bfdata{X}
	\right],
\end{equation}
such that
\begin{equation}\label{eq:diffh}
	\frac{\delta \mathcal{L}[\mathbf{f}]}{\delta \mathbf{f^\dagger}}
	=
	\langle \mathbfcal{H} \rangle - \mathbfcal{H}
	=
	0.
\end{equation}
The expectation value $\langle \mathbfcal{H} \rangle$ corresponds to a ``mean field'',
which comes from the  derivative of the determinant term. It can conveniently be thought of as a data simulation average.%
\footnote{From the identity $\mathrm{Tr}\mathbf{A} = \langle \mathbf{x}^\dagger \mathbf{A} \mathbf{C}^{-1} \mathbf{x} \rangle$, which holds for a random vector $\mathbf{x}$ with $\langle \mathbf{x} \mathbf{x}^\dagger\rangle = \mathbf{C}$ and any (square) matrix $\mathbf{A}$, we have
\begin{equation*}
	2 \langle \mathbfcal{H} \rangle =
	\left\langle
	\bfdata{X}^\dagger
	\Big(\mathbf{C^{\hat{X}\hat{X}}}\Big)^{-1}
	\frac{\mathbf{\delta C^{\hat{X}\hat{X}}}}{\delta \mathbf{f^\dagger}}
	\Big(\mathbf{C^{\hat{X}\hat{X}}}\Big)^{-1}
	\bfdata{X}
	\right\rangle
	=
	\mathrm{Tr}
	\left[
	\Big(\mathbf{C^{\hat{X}\hat{X}}}\Big)^{-1}
	\frac{\mathbf{\delta C^{\hat{X}\hat{X}}}}{\delta \mathbf{f^\dagger}}
	\right].
\end{equation*}
}
Solutions for $\mathbf{f}$ that satisfy Eq.~\eqref{eq:diffh} can be found iteratively, e.g. with a Newton-Raphson approach. This is particularly efficient for weak sources of anisotropy since $\mathbf{f}\approx\mathbf{0}$, meaning that a single iteration from $\mathbf{f}=\mathbf{0}$ will yield a good approximate solution to the anisotropy. Explicitly, the first iteration produces the estimator
\begin{equation}\label{eq:iter}
	\mathbf{\hat f} = -\left[
	\frac{\delta}{\delta \mathbf{f^\dagger}}
	(\langle \mathbfcal{H} \rangle - \mathbfcal{H})^\dagger
	\Big|_0^\dagger
	\right]^{-1}
	(\langle \mathbfcal{H}_0 \rangle - \mathbfcal{H}_0),
\end{equation}
where the lower index ``0'' indicates evaluation with $\mathbf{f}=\mathbf{0}$, corresponding to the initial step. The functional derivative in Eq.~\eqref{eq:iter} can be approximated with the Fisher information
\begin{equation}
	\mathbfcal{F}_0 =
	\left\langle
	\frac{\delta}{\delta \mathbf{f^\dagger}}
	(\langle \mathbfcal{H}_0 \rangle - \mathbfcal{H}_0)
	\right\rangle,
\end{equation}
which provides a natural form of normalization for the likelihood based estimator. The normalized quadratic estimator that we use then takes the form
\begin{equation}\label{eq:qe_basic}
	\mathbf{\hat{f}} =
	\mathbfcal{F}^{-1}_0(\mathbfcal{H}_0 - \langle\mathbfcal{H}_0\rangle),
\end{equation}
where
\begin{equation}\label{eq:H0_term}
	\mathbfcal{H}_0 =
	\frac{1}{2}
	\mathbf{\mathbf{\bar{X}}}^\dagger
	\left[
	\frac{\delta \mathbf{C^{X X}}}{\delta \mathbf{f}^\dagger}
	\right]_0
	\mathbf{\mathbf{\bar{X}}},
\end{equation}
and we have introduced the inverse (co-)variance filtered fields, defined as $\mathbf{\mathbf{\bar{X}}}\equiv \mathbfcal{B}^\dagger(\mathbf{C^{\hat{X}\hat{X}}})^{-1}\big|_0 \bfdata{X} $. An analytic expression for the Fisher normalization is in general non-trivial, so in practice we consider the isotropic full-sky limit of the reconstruction to obtain a diagonal simplification that we then correct for with a Monte Carlo correction. We implement a normalization that is consistent\footnote{The normalization differs in the case of the joint estimators as defined in~Eq.\eqref{eq:symm_fisher}, but is identical otherwise.}
with that given in Refs.~\cite{Hanson:2009gu, Pearson:2012ba}.

The covariance $\mathbf{C^{{X}{X}}}$ is (in the general case) a $3\times 3$ block matrix, whose elements $\mathbf{C}^{s_1 s_2} = \langle {}_{s_1} \vX \ {}_{s_2} \vX^\dagger \rangle$ represent the covariance between the spin fields with fixed anisotropy. We can rewrite the product of matrices as the sum over spin-indices to obtain
\begin{equation}\label{eq:spin_qe_sum}
	\mathbfcal{H}_0 =
	\frac{1}{2}
	\sum_{{s_1}, {s_2}} {}_{s_1} \mathbf{\bar{X}}^\dagger
	\left[\frac{\delta \mathbf{C}^{{s_1}{s_2}}}{\delta \mathbf{f}^\dagger}\right]_0
	{}_{s_2} \mathbf{\bar{X}}.
\end{equation}
We calculate (up to linear order in modulation) the derivative of the covariance to be
\begin{equation}\label{eq:dcov_df}
	\left[
	\frac{\delta \mathrm{C}^{{s_1}{s_2}}(\vnhat_1, \vnhat_2)}{\delta f(\vnhat)}
	\right]
	=
	\langle \rmspincomp{s_1}{X}(\vnhat_1)\ \rmspincomp{s_2}{X}^*(\vnhat_2)\rangle(\delta(\vnhat - \vnhat_1) + \delta(\vnhat - \vnhat_2)),
\end{equation}
where $\delta(\vnhat - \vnhat_i)$ appearing on the right-hand side result from locality of the modulation. By expanding $\rmspincomp{s}{X}(\vnhat)$ into spherical harmonics Eq.~\eqref{eq:dcov_df} can be written in terms of the fiducial theoretical power spectra associated with the spin fields
\begin{equation}
	\langle \rmspincomp{s_1}{X}(\vnhat_1)\ \rmspincomp{s_2}{X}^*(\vnhat_2)\rangle
	=
	\sum_{\ell_1, m_1}\sum_{\ell_2, m_2}
	\mathcal{B}_{\ell_1} \mathcal{B}_{\ell_2}
	\ {}_{s_1}Y_{\ell_1 m_1}(\vnhat_1) \ {}_{s_2}Y_{\ell_2 m_2}^*(\vnhat_2)
	\langle
	{}_{s_1} X_{\ell_1 m_1}(\vnhat_1) \
	{}_{s_2} X_{\ell_2 m_2}^*(\vnhat_2)
	\rangle,
\end{equation}
where $\mathcal{B}_\ell$ is the harmonic space representation of the instrumental beam and pixel window $\mathbfcal{B}$.
The quadratic building block for the modulation estimator is then
\begin{equation} \label{eq:modquadratic}
	{}_\mathcal{S} \mathcal{H}_0^{X X} (\vnhat)
	=
	\sum_{s\in\mathcal{S}} {}_s\bar{X}(\vnhat)^* {}_sX^\mathrm{WF}(\vnhat),
\end{equation}
where $\mathcal{S}$ is the set of CMB field spin indices to sum over.
In practice, we implement a version of conjugate-gradient descent to obtain an approximate iterative solution for the Wiener-filtered (WF) fields ${\vX}^\mathrm{WF} = \mC\mathbf{\mathbf{\bar{X}}}\equiv \mC\mathbfcal{B}^\dagger(\mathbf{C^{\hat{X}\hat{X}}})^{-1}\big|_0 \bfdata{X}$, where $\mathbf{C}$ is the covariance of fiducial power spectra, for both data and the simulations~\cite{Smith:2007rg,Aghanim:2018oex}. The inverse-variance filtered (IVF) fields are simply related to the WF fields by dividing the matrix of fiducial spectra $\mathbf{C}$. The quadratic estimator remains unbiased, though slightly suboptimal~\cite{Maniyar:2021msb}, if an approximate version of $\mathbf{C^{\hat{X}\hat{X}}}$ is used, as long as the normalization is computed consistently.
For simplicity, we take $C^{TE}=0$ in the theory part of the covariance $\mathbf{C^{\hat{X}\hat{X}}}$ to allow separate filtering of temperature and polarization maps, and approximate the noise as isotropic over the unmasked area. The inverse noise is taken to be zero in the masked regions, so that masked pixels do not contribute to the WF maps.
The modulation estimator does respond to $B$ modes, but for \planck\ they contribute negligibly to the $\taunl$ constraint.\footnote{A scalar primordial non-Gaussianity signal should not produce $B$ modes, but a modulation does. In the ultra-squeezed limit in which the modulation approximation is exact, the $B$-mode contribution also vanishes. Including the $B$ modes is correct for modulation reconstruction, but for futuristic measurements where the $B$-mode contribution may become non-negligible due to low delensed cosmic variance, the $B$ mode contribution should strictly be dropped when using the modulation model as an approximation to constrain $\taunl$. }
 The equivalent harmonic space representation of~\eqref{eq:modquadratic} can be directly obtained via the spin-zero harmonic transformation
\begin{equation}\label{eq:harmonic_mod_quadratic}
	{}_\mathcal{S} \mathcal{H}_{0, LM}^{X X} = \int \mathrm{d} \vnhat \ {}_\mathcal{S} \mathcal{H}_0^{X X} (\vnhat) _0 Y_{LM}^*(\vnhat).
\end{equation}

By defining the set of spin-indices, $\mathcal{S}$ appearing in Eq.~\eqref{eq:modquadratic} we can construct different estimators. For example, we obtain the (unnormalized) temperature estimator given
in~\cite{Hanson:2009gu} by setting $\mathcal{S}_\mathrm{TT} = \{0\}$
\begin{equation}\label{eq:estimator_tt}
	{}_{\mathcal{S}_\mathrm{TT}} \mathcal{H}_0^{XX}(\vnhat) =
	\bar{{T}}(\vnhat) T^\mathrm{WF}(\vnhat),
\end{equation}
whereas the polarization and minimum variance estimators are defined by $\mathcal{S}_\mathrm{PP} = \{+2, -2\}$ and $\mathcal{S}_\mathrm{MV} = \{0, +2, -2\}$ respectively. These latter estimators have simple expressions analogous to Eq.~\eqref{eq:estimator_tt}. Each estimator can equivalently be expressed in terms of the CMB harmonics $\{T_{lm}, E_{lm}, B_{lm}\}$, though appear more cumbersome. 

The quadratic estimators can reconstruct the modulation field up to the limits of instrumental noise and the cosmic variance of the CMB fields themselves. These Gaussian sources produce a reconstruction noise, $N_L^{(0)}$, which corresponds directly to the disconnected trispectrum contributions discussed in~\ref{section:squeezed_taunl}, and should be subtracted from any modulation field power spectrum estimate. This can be seen by computing the covariance between two quadratic estimators, which will result in Gaussian terms, composed of products of noise and (theoretical) CMB power spectra. Under idealized conditions, this reconstruction noise is statistically isotropic and equivalent to the isotropic Fisher normalization, such that it is straightforward to account for. We normalize using the full-sky analytic result,
but correct our final spectra by a Monte Carlo normalization to account for inaccuracies in the analytic result. The reconstruction noise bias and mean field are also calculated from simulations, as described in the following section. Using harmonic representation of the full-sky analytic Fisher normalization gives us the normalized quadratic estimator
\begin{equation}\label{eq:normed_modquadratic}
	{}_\mathcal{S}\hat{f}_{LM}^{XX} =
	({}_\mathcal{S} \mathcal{F}^{XX}_{0, L})^{-1}
	{}_\mathcal{S} \mathcal{H}_{0, LM}^{X X},
\end{equation}
where we have appended indices to the normalization factor in order to keep track of the estimator properties, which will alter the required normalization in general.

We can generalize the quadratic estimator in the Eq.~\eqref{eq:modquadratic} to a ``joint estimator'', which combines information from pairs of maps which may correspond to e.g. distinct observations of the CMB. For map pairs $X(\vnhat), Z(\vnhat)$ we define the unnormalized estimator analogous to~\eqref{eq:modquadratic}
\begin{equation} \label{eq:joint_modquadratic}
	{}_\mathcal{S} \mathcal{H}_0^{X Z} (\vnhat)
	=
	\frac{1}{2}
	\sum_{s \in \mathcal{S}}	
	\left[
	{}_s\bar{Z}^*(\vnhat) {}_s X^\mathrm{WF}(\vnhat) +
	{}_s\bar{X}^*(\vnhat) {}_s Z^\mathrm{WF}(\vnhat)
	\right],
\end{equation}
where the symmetrized mean accounts for the fact that in general the quadratic building block is not symmetric under exchange $X\leftrightarrow Z$. Similarly, we write the symmetrized normalization as
\begin{equation}\label{eq:symm_fisher}
	{}_\mathcal{S} \mathcal{A}^{XZ}_{L}
	=
	\frac{1}{2}
	\Big(
	{}_\mathcal{S} \mathcal{F}^{X Z}_{0, L} +
	{}_\mathcal{S} \mathcal{F}^{Z X}_{0, L}
	\Big),
\end{equation}
to obtain
\begin{equation}\label{eq:normed_qe}
	\flmhat =
	({}_\mathcal{S} \mathcal{A}^{XZ}_{L})^{-1}
	{}_\mathcal{S} \mathcal{H}_{0, LM}^{X Z}.
\end{equation}
Note that if $X=Z$ these definitions automatically recover the former result of Eq.~\eqref{eq:normed_modquadratic}. In the remainder of this paper we will use power spectrum estimates of the modulation field computed from the normalized quadratic estimator(s) described in this section.

%% file: sections/theory_and_methodology/subsection_power_spectrum_estimator.tex
\subsection{Power spectrum estimator}\label{section:mf_rdn0}
Power spectrum estimates of the modulation field are computed from the cross-spectrum of pairs of quadratic estimators $\flmhatone$ and $\flmhattwo$. Each of these components require mean field subtraction as given in Eq.~\eqref{eq:diffh}, which in practice is calculated as the average quadratic estimate over simulations with $f=0$. The mean field serves as a map-level correction to the data that accounts for the main sources of statistical anisotropy that are not primordial. These include Doppler aberration and modulation, anisotropic noise and beam anisotropy, as well as estimator response to masked regions of the sky. The accuracy to which the mean field can be estimated is limited by the number and fidelity of the simulations.
The Doppler-induced effects are important for $\taunl$ since most of their signal is at very low $L$, but have a fixed known direction such that they are expected to be accounted for accurately by the simulations.

Monte Carlo noise on the mean field limits accuracy of the reconstruction, but can on average be nulled at the power spectrum level by ensuring the mean field subtracted from the data reconstruction on each leg has independent Monte Carlo noise.
Since realizations of CMB and instrumental noise are uncorrelated from simulation-to-simulation, this can be achieved straightforwardly by constructing two separate mean field estimates from independent sets of simulations.
The mean field (on each component $i\in\{1, 2\}$) is
\begin{equation}\label{eq:mean_fields}
	{}_{\mathcal{S}_i} f_{LM}^{X_i Z_i, \mathrm{MF}} =
	\big \langle {}_{\mathcal{S}_i} \hat{f}^{X_i Z_i}_{LM} \big \rangle_{\mathrm{sims}_i}
\end{equation}
such that the leading contribution to the modulation power spectrum is given by\footnote{%
Our use of ``$\supset$'' in this section implies that whatever appears on the right-hand side is a partial contribution to the full quantity given on the left; which is added to linearly.
}
\begin{equation}\label{eq:cl_mf_subracted}
	\clmodfbaronetwo \supset \frac{1}{(2L + 1)\fsky}\sum_{M} \left( \flmhatone -\mfone\right) \left(\flmhattwo - \mftwo\right)^*.
\end{equation}

If we assume that the simulations reliably capture the expected lensing signal\footnote{The simulations have independent random lensing realizations, so lensing produces no mean field on average, but does have a trispectrum. In this paper we do not attempt a joint reconstruction of the lensing and the modulation.} the estimator can then be made unbiased by simply subtracting the mean over simulations with zero modulation $\clmodfbaronetwo  - \langle \clmodfbaronetwo \rangle_{\modf=0}$. However, we can instead use the optimal perturbative trispectrum estimator that does better than this, by correcting for the realization of the Gaussian power in the data map compared to the mean in the simulations~\cite{Regan:2010cn,Hanson:2010rp,Ade:2015zua,Namikawa2013}. This realization-dependent correction, $\rdnzero$, reduces the variance of the estimator, and also makes it perturbatively self-correcting to misestimates of the Gaussian power in the simulations. In the idealized full sky case this can be done by adding the linear correction
\begin{equation}
	\clmodfbaronetwo \supset
	\langle \clmodfbaronetwo \rangle_{\modf=0}
	- \sum_{ab,l}
	\frac{\partial\langle \clmodfbaronetwo \rangle_{\modf=0}}{\partial \bar{C}_{\ell,\rm{expt}}^{ab}}
	\left( \bar{C}_{\ell,\rm{expt}}^{ab} - \bar{C}_{\ell,\rm{expt, fid}}^{ab}\right),
\end{equation}
where $\bar{C}_{\ell,\rm{expt}}^{ab}$ is the CMB power spectrum measured on the data and $\bar{C}_{\ell,\rm{expt}, fid}^{ab}$ the fiducial model of the data power spectrum (i.e. lensed power spectrum + isotropic noise).
The dominant part of the derivative correction comes from the response of $\nzero$ to the power, so we can approximate $\langle \clmodfbaronetwo \rangle_{\modf=0}\approx \nzero$ in that term; the first term depends on $\nzero$ and the lensing 4-point contribution to the modulation power.

Accounting for these bias corrections, our estimator for the modulation power spectrum is given by
\begin{equation}\label{eq:uncal_total_cl}
	\clmodfbaronetwo
	=
	\frac{1}{(2L + 1)\fsky}\sum_{M} \left( \flmhatone -\mfone\right) \left(\flmhattwo - \mftwo\right)^*
	- \deltalens  - \rdnzero,
\end{equation}
where $\deltalens = \langle \clmodfbaronetwo \rangle_{\modf=0} - \nzero$ is the connected lensing contribution.  In the general non-full-sky case of power spectra based on four input maps $\{X_1, Z_1, X_2, Z_2\}$ the explicit expressions for these terms are:
\begin{equation}\label{eq:lensbias}
	\begin{split}
		\deltalens =
  		\Big\langle
        \bar{C}_L
        ^{
        \mathcal{S}_1
        \field{X_1}{i}
        \field{Z_1}{i}, \
        \mathcal{S}_2
        \field{X_2}{i}
        \field{Z_2}{i}
        }
        -\Big(
        \bar{C}_L
        ^{
        \mathcal{S}_1
        \field{X_1}{i}
        \field{Z_1}{j}, \
        \mathcal{S}_2
        \field{X_2}{j}
        \field{Z_2}{i}
        }
        + \bar{C}_L
        ^{
        \mathcal{S}_1
        \field{X_1}{i}
        \field{Z_1}{j}, \
        \mathcal{S}_2
        \field{X_2}{i}
        \field{Z_2}{j}
        }
        \Big)
        \Big\rangle_{i \neq j, \modf=0}
	\end{split}
\end{equation}
and
\begin{equation}\label{eq:rdn0}
    \begin{split}
       \rdnzerobar =\Big\langle
       &\bar{C}_L
       ^{
        \mathcal{S}_1
       \field{X_1}{i}
       \field{Z_1}{d}, \
        \mathcal{S}_2
       \field{X_2}{i}
       \field{Z_2}{d}
       }
       + \bar{C}_L
       ^{
        \mathcal{S}_1
       \field{X_1}{i}
       \field{Z_1}{d}, \
        \mathcal{S}_2
       \field{X_2}{d}
       \field{Z_2}{i}
       }
       +  \bar{C}_L
       ^{
        \mathcal{S}_1
       \field{X_1}{d}
       \field{Z_1}{i}, \
        \mathcal{S}_2
       \field{X_2}{i}
       \field{Z_2}{d}
       }\\
       + &\bar{C}_L
       ^{
        \mathcal{S}_1
       \field{X_1}{d}
       \field{Z_1}{i}, \
        \mathcal{S}_2
       \field{X_2}{d}
       \field{Z_2}{i}
       }
       - \Big(\bar{C}_L
       ^{
        \mathcal{S}_1
       \field{X_1}{i}
       \field{Z_1}{j}, \
        \mathcal{S}_2
       \field{X_2}{j}
       \field{Z_2}{i}
       }
       + \bar{C}_L
       ^{
        \mathcal{S}_1
       \field{X_1}{i}
       \field{Z_1}{j}, \
        \mathcal{S}_2
       \field{X_2}{i}
       \field{Z_2}{j}
       }\Big)
       \Big\rangle_{i \neq j, \modf=0}
    \end{split}
\end{equation}
where the indices $(i)$ and $(j)$ imply the maps are Monte Carlo simulations, and $(d)$ the data. In practice we implement the convention $j = i + 1$ to obtain uncorrelated Monte Carlo pairs, though in principle further permutations of indices could lower the Monte Carlo noise even further. The power spectra have the same definition as in Eq.~\eqref{eq:cl_mf_subracted} other than the mean field term, which is set to zero. The exception to this is the first term appearing in Eq.~\eqref{eq:lensbias} when each map carries the same index. This term includes lensing 4-point information and will play a further role in our likelihood analysis, it is computed from individual Monte Carlo simulations as
\begin{equation}\label{eq:mcn0}
	\mcnzerobar = \big\langle
        \bar{C}_L
        ^{
        \mathcal{S}_1
        \field{X_1}{i}
        \field{Z_1}{i}, \
        \mathcal{S}_2
        \field{X_2}{i}
        \field{Z_2}{i}
        }
    \big\rangle_{\modf=0}.
\end{equation}

Finally we introduce a calibration factor $k_L$ to form the final estimate
\begin{equation}
	\clmodfhatonetwo = k_L \clmodfbaronetwo.
\end{equation}
The calibration factor is derived empirically from with-signal simulations such that the ensemble average reconstruction spectra match the theoretical input
$C_L^{f, \mathrm{Theory}} = k_L \langle \clmodfbaronetwo \rangle_{\rm sims}$. Note that when the calibration factor multiplies the reconstruction noise spectra we apply the same convention; e.g. $\mcnzerohat = k_L \mcnzerobar$.

The calibration corresponds to an $\mathcal{O}(1)$ multiplicative correction after appropriate binning and smoothing is applied, though is slightly scale-dependent. The dipole receives the largest correction corresponding to the greatest power loss after masking the sky, and towards smaller scales the power is typically reduced by up to $15\%$.
We assess the calibration procedure in more detail in Section~\ref{sec:pipeline_validation}, and have confirmed that it leads to more accurate $\taunl$ estimates, and does not lead to a false detection when applied to without-signal reconstructions. In the remainder of this paper, spectra that are uncalibrated or calibrated will be indicated with a bar or hat, i.e. $\bar{C}_L$ or $\hat{C}_L$ respectively.

\begin{figure}
	\includegraphics[width=0.75\textwidth]{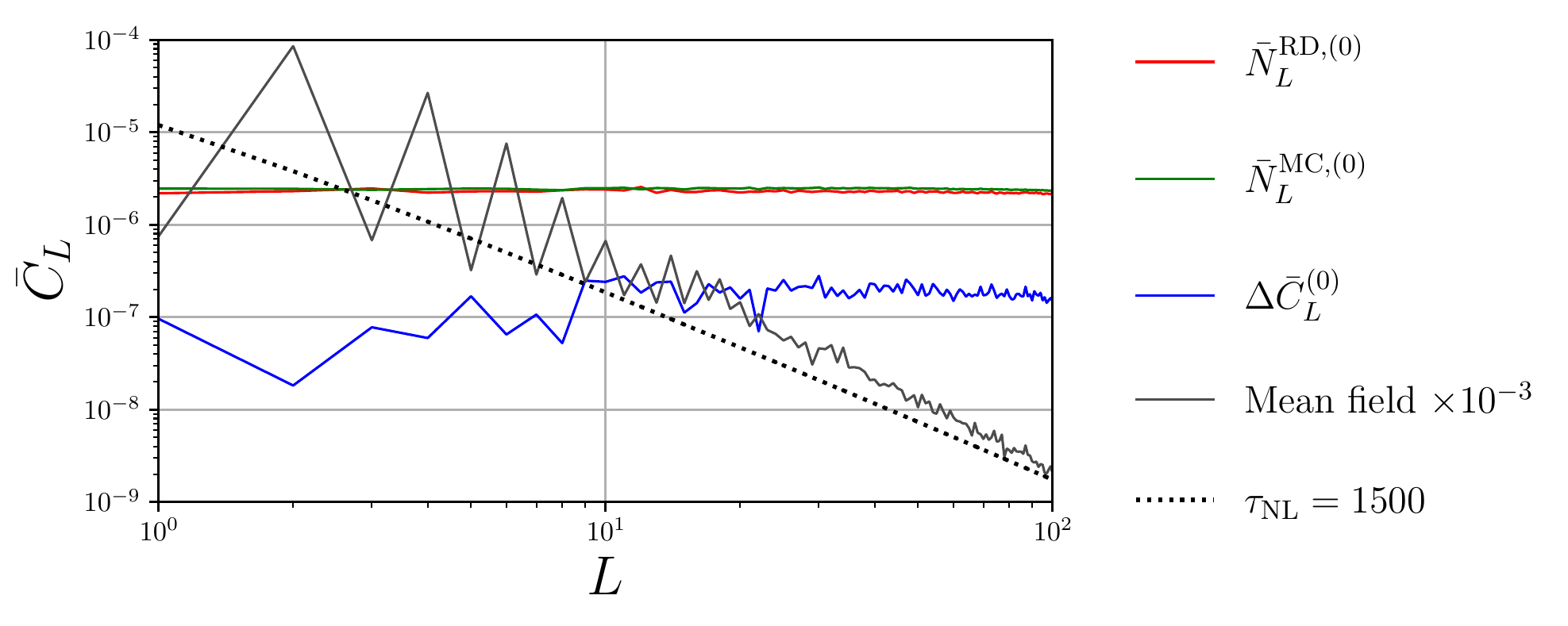}
	\caption{Power spectrum of the modulation mean field, reconstruction noise biases, and lensing bias, for the \npipe (\texttt{A}, \texttt{B}) analysis. The mean field power is computed as the cross-spectrum of the independent mean field components given in Eq.~\eqref{eq:mean_fields}. For comparison, the dotted line shows the modulation power expected if $\taunl=1500$; i.e. $C_L^\modf = 1500 C_L^{\zeta_*}$.
}
	\label{fig:npipe_bias}
\end{figure}

%% file: sections/theory_and_methodology/subsection_likelihood.tex
\subsection{$\taunl$ estimation and likelihood analysis}\label{sec:taunl_likelihood}
The simplest estimate that we can construct for $\taunl$ corresponds to the modulation power spectrum at a fixed scale $L$, normalized by the curvature power spectrum at recombination
\begin{equation}\label{eq:simple_estimator}
	\hat{\tau}_\mathrm{NL}(L) = \frac{\clmodfhat}{C_L^{\zeta_*}},
\end{equation}
where we assume that we know $C_L^{\zeta_*}$ theoretically; and can be readily computed with e.g. \href{http://camb.info}{\texttt{CAMB}}. In general $\clmodfhat = \clmodfhatonetwo$ (i.e. is the full modulation power spectrum estimate) but we have omitted indices for convenience. The calibrated estimates are related to the uncalibrated estimates multiplicatively $\clmodfhat = k_L \clmodfbar$, such that $\hat{\tau}_\mathrm{NL}(L) = k_L\bar{\tau}_\mathrm{NL}(L)$. We adopt the same bar/hat convention for all quantities in our likelihood analysis to make this distinction clear; including the covariance and Gaussian noise bias.
Estimates at each $L$ can be combined using an inverse variance weighting. For small signals, an approximately scale-invariant primordial spectrum, and approximating the reconstruction noise as white over the modulation multipole range $L_\mathrm{min}\leq L \leq L_\mathrm{max}$ that is used, this gives the approximately-optimal combined estimator~\cite{Pearson:2012ba}
\begin{equation} \label{eq:finite_estimator}
	\hat{\tau}_\mathrm{NL} =
	\frac{1}{R}
	\sum_{L=L_\mathrm{min}}^{L_\mathrm{max}}
	\frac{2L + 1}{L^2 (L + 1)^2}
	\frac{\hat{C}_L^\modf}{C_L^{\zeta_*}},
\end{equation}
where the normalization $R$ accounts for the finite multipole range, $R = {L_\mathrm{min}}^{-2} - (L_\mathrm{max} + 1)^{-2}$. Regardless of whether or not there is in fact a primordial modulation field, the reconstruction is expected to produce an approximately Gaussian random field on large scales. In the null case, the reconstruction should correspond to a realization of the disconnected Gaussian noise bias, $N_L^{(0)}$, with an approximately white power spectrum. If there is a primordial modulation, this too is likely to correspond to an approximately Gaussian random field (e.g., if it is generated by quantum fluctuations early in inflation). If it is of $\taunl$ form, the spectrum  would peak on large scales if the signal-to-noise of the modulation is sufficiently large to be distinguishable from Gaussian background. We therefore expect that the estimators formed as a sum over $\hat{\tau}_\mathrm{NL}(L)$ are approximately $\chi^2$-distributed. Since the signal peaks at low multipoles, this is a very skewed distribution, with $95\%$ of the signal in $\htaunl$ at $L\leq 4$~\cite{Pearson:2012ba}.

To constrain the true value of $\taunl$ given the high level of realization dependence, we use an approximate model for the posterior distribution of the set of $\hat{\tau}_\mathrm{NL}$ estimates. As pointed out in Ref.~\cite{Planck:2013wtn}, the statistical properties of a $\taunl$ signal are analogous to those assumed when estimating the statistics of the CMB temperature field. We largely borrow from this ``single-field'' procedure and construct an approximate log-likelihood function on the cut-sky as described in Ref.~\cite{Hamimeche:2008ai}
\begin{equation}\label{eq:log_like}
	-2 \ln P
	\left(\{\taunl(L)\} | \{\htaunl(L)\}\right)
	\approx
	\sum_{L L'}
	\Big[ g(\hat{x}(L)) \hat{N}^{\mathrm{MC},(0)}_{\taunl}(L)\Big]
	\Big[\hat{M}^{-1}\Big]_{L L'}
	\Big[ \hat{N}^{\mathrm{MC},(0)}_{\taunl}(L') g(\hat{x}(L'))\Big]+\text{const,}
\end{equation}
using a restricted multipole range $L_\mathrm{min}\leq L \leq L_\mathrm{max}$. In the above $\htaunl(L)$ is defined in Eq.~\eqref{eq:simple_estimator}, and $\hat{N}_\taunl^{\mathrm{MC},(0)}(L)$ is of the same form but using the reconstruction noise power spectrum in place of the modulation power spectrum (and therefore defines the Gaussian noise bias on $\taunl$).
$\hat{M}$ is the covariance matrix of $\htaunl(L)$ estimates inferred from without-signal simulation reconstructions, and is truncated to the prescribed multipole interval before inversion.
To better account for the error incurred by the inverse-covariance due to the finite number of simulations, we rescale the inverse by a constant (Hartlap factor~\cite{Hartlap:2006kj}) which leads to a slight broadening on the final posterior distribution. Similarly, we account for potential Monte Carlo errors (i.e. error on the mean field and reconstruction noise bias) by including the additive corrections to the diagonal of the covariance~\cite{Aghanim:2018oex}
\begin{equation}
	\sigma_{\mathrm{MC}, L}^2 =
	\left(
	\frac{2}{N_\mathrm{MF}} + \frac{9}{N_\mathrm{bias}}
	\right)
	\frac{2 (N_\taunl^{(0)})^2}{(2 L + 1)\fsky},
\end{equation}
where $N_\mathrm{MF}$ and $N_\mathrm{bias}$ are the number of simulations used to compute the mean field and realization dependent noise bias respectively.
The likelihood further relies on the functions
\begin{equation}\label{eq:g_x_log_like}
	g(x) =
	\mathrm{sign}(x - 1)
	\sqrt{2 (x - \ln(x) - 1)},
	\quad \mathrm{where} \quad \hat{x}(L) =
	\frac{\htaunl(L) + \hat{N}_\taunl^{\mathrm{RD}, (0)}(L)}{\taunl(L) + \hat{N}_\taunl^{\mathrm{RD}, (0)}(L)},
\end{equation}
where $\hat{N}_\taunl^{\mathrm{RD},(0)}(L)=k_L \rdnzerobar/C_L^{\zeta_*}$ accounts for realization dependent Gaussian noise fluctuations in the data.%
\footnote{An exception to this may be ``TTxPP'' estimators. In this case, the resulting $\rdnzero$ spectra have a small amplitude, which may be negative for some multipoles. For small $\taunl$ the logarithm in Eq.~\eqref{eq:g_x_log_like} will then become undefined. Alternative approximations to the reconstruction noise should then instead be used, e.g. $N^{(0)}_\taunl(L) \simeq \sqrt{(2L + 1)\fsky M_{L L} / 2}$ which is non-negative by definition.}
Although we reconstruct the modulation field with general $L$ dependence, for constraining $\taunl$ we follow the standard definition by assuming no scale dependence in $\taunl$ itself; $\taunl(L)=\taunl$.

Once we have obtained our posterior distribution of $\taunl$ estimates, it is straightforward to construct constraints on $\taunl$ by numerically integrating the distribution from $\taunl=0$ up until some desired limit. Recall that $\taunl$ is positive definite by definition in the context of squeezed primordial non-Gaussianity.

%% file: sections/data_and_sims/data_and_sims.tex
\section{Data and simulations}\label{sec:data_and_sims}

The nominal \Planck\ 2013 results placed constraints on \taunl using the 143 and 217 GHz intensity maps to reconstruct the modulation field. The foreground signal was partly mitigated by projecting a dust template from higher frequencies, but the \Planck\ 2013 analysis still showed evidence of important residuals.
There is also a large spatial variation of the noise, which can mimic the effect of spatially varying power from $\taunl$ if it cannot be simulated and corrected for accurately.  This noise modulation can be mitigated by correlating different maps with independent noise;
in \Planck\ 2013 a large quadrupole was substantially reduced by performing the joint (143, 217) GHz reconstruction. The joint reconstruction is obtained by direct application of the estimators discussed in section~\ref{sec:modulation_estimator}, where one copy of the inverse filtered harmonics (e.g., as given in Eq.~\eqref{eq:estimator_tt}) is replaced with those corresponding to a different map. More generally, the noise from observations at different times and different detectors is also approximately uncorrelated, and we can also use joint reconstruction between different data subsets to reduce sensitivity to the spatially-varying noise power.

Advancing from the 2013 to 2018, the data products are derived from a longer observation period via repeated scans of the sky, leading to a lower effective noise in all frequency channels. In addition to this, the release of polarization data enables us to expand the range of estimators at our disposal; from temperature only (TT), to polarization (PP) and minimum variance (MV) combinations, as defined in Section~\ref{sec:modulation_estimator}.
PP estimators typically offer weak constraining power in comparison to TT and MV, since the small-scale signal is dominated by noise for \Planck. 
However, the MV estimate (which combines information from polarization with temperature), by definition reduces the reconstruction noise when compared to TT and PP (if implemented optimally), and we use it for our best constraints. 
Other cross-power spectrum estimators can be built by considering the cross between estimators of different types, such as ``TT$\times$PP''. Whilst this estimator has low reconstruction noise due to the small noise cross-power spectrum between the temperature and polarization fields, this comes at the expense of a larger estimator variance such that the resulting spectra provide relatively weak constraints.

To see the impact of the changes since the 2013 result, we first produce constraints from updated 143 GHz and 217
GHz temperature-only data, to illustrate improvements associated with lowering the effective noise, whilst remaining
subject to the same frequency-dependent signals as in 2013. Secondly, we then use the component-separated \smica\
maps in temperature and polarization to reduce foreground contamination, using the joint reconstruction between
observations in first and second half of the observational period (\hmone, \hmtwo) to mitigate noise modelling uncertainty.
Comparison between temperature reconstruction and the minimum-variance joint constraint with polarization (MV)
shows the impact of the polarization signal, while also have different dependence on foreground residuals (which are
expected to be lower in polarization on small scales). Finally, we use the latest \npipe data and simulation set, which
have (in most respects) higher-fidelity noise-modelling and signal processing. We use the joint MV reconstruction
between \npipe time-period data splits \splita and \splitb to obtain our baseline results.

In the remainder of this section, we provide further technical details regarding the processing of the data and simulations that propagate into our estimators, and finally the likelihood analysis.

\input{sections/data_and_sims/subsection_simulations}

%% file: sections/data_and_sims/subsection_simulations.tex
\subsection{Simulations and Corrections}

The Full Focal Plane (\texttt{FFP}) simulations~\cite{Ade:2015via,Aghanim:2018fcm} aim to replicate the statistics of the observed CMB sky. The end products from the map-making process are a set of lensed scalar realizations of the cosmological CMB along with simulated noise + residual systematics. The CMB simulations rely on fiducial power spectra from which scalar unlensed CMB realizations are constructed. The unlensed CMBs are then lensed by Gaussian realizations of a lensing potential that are independent between simulations. The input cosmological parameters used in the process of generating and subsequently lensing the scalar maps are based on \Planck\ public release (PR) data. For example, in the case of \ffpten,
which are the set of CMB and noise simulations that accompany the \Planck\ PR3 data,
the CMB simulations assume a base $\Lambda$CDM cosmology with parameters inferred from previous PR2~\cite{Ade:2015xua} data.
Note that the PR3 data are also referred to as the ``\dxtw'' data - the latter of which usually appears in technical documentation - and we use these interchangeably throughout this work.
The simulated CMB signal contains additional frequency-dependent sources. These include Rayleigh scattering, a second order dipole-induced quadrupole~\cite{Kamionkowski:2002nd}, and a Doppler modulation and (frequency-independent) aberration~\cite{Aghanim:2013suk}. Once the lensed scalar maps are generated, they are convolved with \texttt{FEBeCoP}~\cite{Mitra:2010rt}, which accounts for realistic beaming and instrumental pointing.

In order to generate the accompanying \ffpten noise maps a fiducial realization of the scalar CMB is first constructed. The process further relies on sophisticated foreground modelling, via the Planck Sky Model~\cite{PSM:2013}, which contains galactic thermal dust, spinning dust, synchrotron, free-free scattering and CO line emission based on data-based models, as well as random realizations of CIB and SZ that are (unrealistically) uncorrelated to the fiducial CMB lensing potential realization. The nominal input sky map in conjunction with the fiducial CMB realization is then processed to include systematic effects such as detector pointing, and known point sources~\cite{Reinecke:2005fv}.
From this point, independent realizations of noise are added to mimic the observational data stream~\cite{Planck:2018lkk}. Finally, by subtracting out the fiducial CMB and foregrounds, the noise (+ systematics) simulations are obtained.

In addition to the \dxtw data and \ffpten simulations, in this work we also make use of the PR4  \npipe\ data~\cite{Planck:2020olo} and its corresponding simulations.
At the level of the data, there are improvements in the polarization calibration procedure, and additional data ($\sim 8\%$) from the (previously neglected) satellite repointing periods, leading to a slightly reduced noise level. Enhanced data processing further accounts for systematic effects by building templates in the time-ordered data stream. These template amplitudes can be inferred via marginalization over the sky, and lead to iteratively ``cleaned'' maps as the template amplitudes converge to zero.  With respect to the simulations, the lensed scalar CMB remains unchanged from \ffpten, but the end-to-end simulations emulate many features from the data processing pipeline, such as the removal of systematics in time ordered information previously discussed. Finally, the \npipe\ processing allows for the construction of maximally independent detector-set split maps, which are referred to as the \splita and \splitb splits. These should in principle provide improved noise decorrelation with respect to one another than that between the half mission splits.

Despite the sophisticated instrumentation model used in \ffpten and \npipe, small discrepancies between the power spectra of the simulations and those observed in the equivalent data will persist. Small spatially varying differences in power could bias our inference of primordial power modulation. Fortunately, the realization-dependent disconnected noise correction ($\rdnzero$) that we apply automatically corrects for the difference between the true and simulation covariance to leading order, making the estimators relatively robust. However, for higher-order differences to be small, we do still need to ensure that the power in the simulations is at least perturbatively close to the power in the data.

We therefore adjust the simulations so that the power spectra of the filtered data and filtered simulations match, targeting consistency to the sub-percent level. In order to do this, we compute the power difference from the WF data and the ensemble average spectrum from WF simulations: $\Delta C_\ell \equiv C_\ell^\text{data} - \langle C_\ell^\text{sims}\rangle$. The corrections are Gaussian and isotropic realizations of power that are added to the simulations and/or data. We defer explicit details of this prescription to Appendix.~\ref{sec:spec_correct} where we explain how we simultaneously correct the auto and cross-spectra between datasets. We illustrate the effectiveness of this corrective procedure for the \dxtw 143/217 GHz data and simulations in~\figref{fig:dcl_npipe_tt}. 

%% file: sections/pipeline_validation/pipeline_validation.tex
\section{Pipeline validation and calibration}\label{sec:pipeline_validation}
We now validate our pipeline by simulating CMB skies modulated with a known $\taunl$ modulation field before adding noise. We check that we recover the modulation spectrum correctly, and that the corresponding posterior is consistent with the input value. 
Since the reconstruction on a single realization is noisy, we check for biases over a set of simulations with the same input $\taunl$. The mean of the reconstructed modulation power spectra should match the input spectrum, and the product of the posteriors should peak close to the input $\taunl$.

We perform these tests using 300 \ffpten \hmone/{\hmtwo} and \npipe \splita/{\splitb} simulations on the cut-sky, using the same masks as in the final data analysis.
In each simulation set, we use common scalar CMB simulations (from \ffpten) modulated with realizations of the modulation field $f(\vnhat)$ with theoretical spectrum $C_L^\modf=\taunl C_L^{\zeta^*} = 1500 C_L^{\zeta^*}$. Key differences in the simulation sets are found in their respective noise maps, as described in Section~\ref{sec:data_and_sims}. After building the modulated simulations, we compute the WF and IVF fields, and then construct ensembles of modulation field estimators from map pairs as given in Eq.~\eqref{eq:normed_qe}. In order to compute the mean field and $N_L^{(0)}$ corrections we repeat this procedure for the same CMB + noise simulations without the modulation signal. Each modulation reconstruction is then mean field and $\mcnzero$ subtracted. We use the same realizations of the noise and unmodulated lensed CMB for the with-signal and without-signal simulations, so that the Monte Carlo and cosmic variance noise largely cancels. This choice enables us to measure any required calibration factors out to small scales, even where the signal is well below the reconstruction noise.

\begin{figure}[h]
	\includegraphics[width=0.9\textwidth]{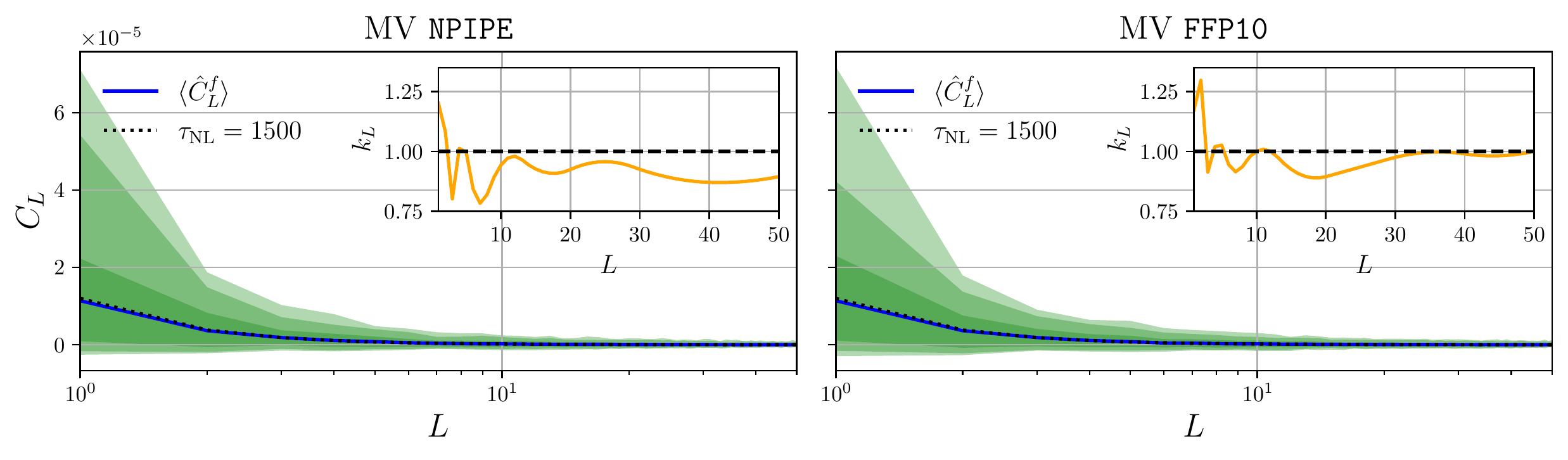}
	\caption{Modulation power spectra from with-signal $\taunl$ simulation minimum variance reconstructions: \textit{Left:} \npipe (\splita, \splitb), \textit{Right:} \ffpten (\hmone, \hmtwo). Dotted line indicates input theory spectrum, and solid blue the mean reconstructed spectrum; with shaded contours indicating the 99, 95 and 68\% empirical C.L. The calibration factor is shown within the inset for each simulation set.
}
	\label{fig:calibrated_cls}
\end{figure}

\begin{figure}[h]
	\includegraphics[width=0.9\textwidth]{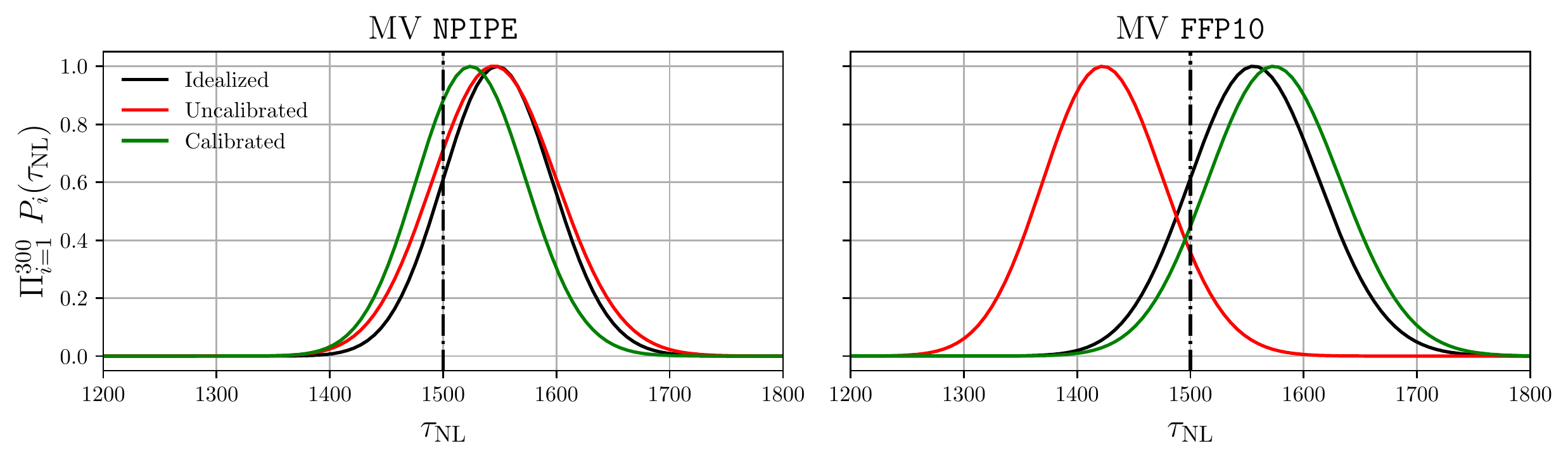}
	\caption{Product of posterior distributions (normalized to one) from 300 with-signal $\taunl$ simulation minimum variance reconstructions: \textit{Left:} \npipe (\splita, \splitb); \textit{Right:} \ffpten (\hmone, \hmtwo). Vertical dash-dotted lines indicates the input simulation value. Solid black posterior assumes perfect full-sky reconstruction of the input modulation power. Red and green lines indicate the uncalibrated and calibrated reconstructions respectively. All posterior distributions are constructed over the multipole range $L\in[1, 50]$.
}
	\label{fig:calibrated_posteriors}
\end{figure}

With the ensemble of reconstructed spectra we compute the calibration factor by comparing their mean against the theoretical input spectrum: $k_L = C_L^{f, \mathrm{Theory}} / \langle \bar{C}_L^f \rangle_\mathrm{sims}$. Since the calibration is empirical (and therefore depends on the details of the simulations, mask and processing) it is not universally defined, and requires expensive calculation. We therefore only compute the calibration for the simulation sets discussed in this section; which include those relevant for our baseline results. In \figref{fig:calibrated_cls} we plot the mean over reconstructed power spectra after $k_L$-correction, and in \figref{fig:calibrated_posteriors} plot the product of the corresponding posterior distributions (normalized by their maximum value).

The $k_L$ factors are plotted within the inset panels of \figref{fig:calibrated_posteriors} for MV reconstructions using the \ffpten and \npipe sets. We further find that the TT calibration factors are approximately equal to those given for MV.
In order to compute $k_L$ we first bin the raw empirical result. For the largest scales of interest $L=1, 2, 3, 4$ we retain unit multipole bins, before incrementally expanding the bin width in multiples of three, weighting by an $L(L+1)$ kernel. The binned solutions are then interpolated to build smooth functions over the smaller scales. This strategy is designed to avoid over-fitting fluctuations in the Monte Carlo noise, whilst modelling the large-scales accurately where the bulk of the modulation signal resides.
At first pass we found that the large-scale calibration leads to over-correction of the posterior estimates, and we therefore decrease the amplitude of $k_L$ by 5\% on these scales.

In \figref{fig:calibrated_posteriors} we plot the calibrated, uncalibrated and ``idealized'' posterior distributions. In the idealized case, we use the (known) realizations of input full-sky modulation power spectra in place of the reconstructions appearing in the estimates of \eqref{eq:g_x_log_like}. This serves as a second reference point in addition to the input $\taunl=1500$ marker, with which the product of posteriors should be consistent. 
Notably this posterior does not peak at the precise fiducial value, which is driven by the cosmic variance and the limited sample size. After calibration, the product of reconstructed simulation posteriors are highly compatible with the product of the idealized posteriors, both of which agree with the theoretical value ($\taunl=1500$) to within $1\sigma$. 
We have further checked that the calibrated posteriors of without-signal simulations are consistent with $\taunl=0$ to the same degree.

%% file: sections/results/results.tex
\section{Results}
\begin{figure}[h]
	\includegraphics[width=0.99\textwidth]{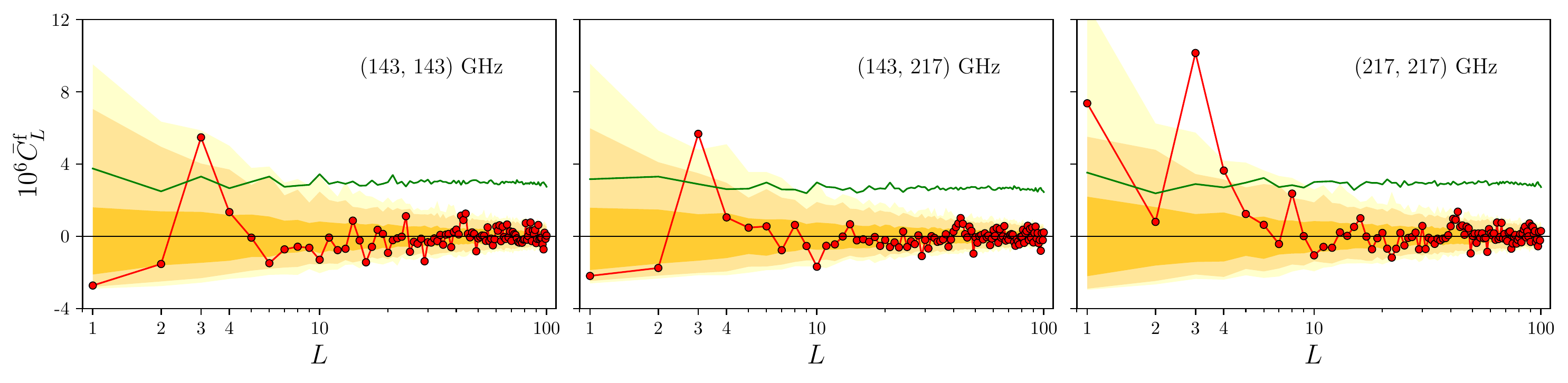}
	\caption{Uncalibrated modulation reconstruction power spectra from \dxtw 143 and 217 GHz data. From left-to-right we plot the auto 143 GHz, joint (143, 217) GHz and auto 217 GHz reconstructions.  Red and green lines indicate the modulation and $\rdnzerobar$ power spectra, and orange bands indicate the 68\%, 95\% and 99\% C.L. of the simulation distribution with no signal.}
	\label{fig:modf_cls_143_217}
\end{figure}

We first report our results for the TT modulation reconstruction obtained from the 143 and 217 GHz channels, which are derived from \texttt{DX12} data and \ffpten\ simulations.
In \figref{fig:modf_cls_143_217} we plot the uncalibrated power spectra for the auto- and joint-reconstructions between these data.
In red we show the modulation field power spectrum after mean field and \rdnzero\ subtraction, the latter of which is given in green for reference.
In orange bands, we plot the 68\%, 95\% and 99\% confidence intervals obtained directly from the equivalent modulation power spectra of without-signal simulations.
This figure is directly comparable to that given in Figure 20 of the \planck\ 2013 non-Gaussianity constraints~\cite{Planck2013-NG}.
Beyond obvious differences in the data and simulations, we note that the methodology presented in this work is slightly different.
In \planck\ 2013 the disconnected noise bias was computed from simulations alone ($\mcnzero$), which is not as robust as the realization-dependent estimate ($\rdnzero$) used in this work.
Moreover in this work we adjust the Gaussian simulation power measured against the data, which makes our $\nzero$ and mean field subtraction more optimal.
When comparing \figref{fig:modf_cls_143_217} to its equivalent in 2013, it should be notated that ``(143, 217) GHz'' in this work equates to ``143$\times$217 GHz'' in the other.
In \planck\ 2013 the 857 GHz intensity map was projected from the 143 and 217 GHz intensity maps whilst obtaining the IVF maps. This projection acted as a template for dusty foreground removal. We project 857 GHz in the same way but now using the \dxtw data.

With this in mind we note that there are differences between the reconstructed spectra, most vividly between the amplitude and sign of the largest scale reconstruction modes.
In \planck\ 2013 there was an $\sim 4 \times$ dipolar power difference between the auto-reconstructions, with the power of the joint reconstruction roughly matching the 143 GHz auto. We find qualitatively similar results; though the difference is $\sim 2\times$ and the sign of our 143 GHz dipole is now negative.
In both analyses, it is likely that the 217 GHz auto signal is strongly contaminated by non-cosmological signal which is not present in the 143 GHz observations. The differences in the dipole sign are partially explained by the realization dependent noise modelling, which slightly boosts the amplitude of $\nzero$ in this work resulting in a more heavily subtracted signal.
The octopole signal similarly has a frequency-dependent amplitude, which in the joint reconstruction approximately match the 143 GHz auto.

The most striking difference is in the quadrupole. Whilst we report that there is very little reconstruction signal in this mode across the auto- and joint-reconstructions, in \planck\ 2013 this dominated the autos before being heavily mitigated in the joint.
The strong frequency-dependent amplitude seen in the \planck\ 2013 analysis already indicated that the bulk of the signal was not from primordial modulation. The difference we observe in this work is most likely explained by better quality (and power-corrected) simulations, which enable us to subtract the frequency-dependent components of the mean field more reliably. Though varying by a lesser extent, we observe a reduced amplitude in the 143 GHz auto, which could be explained in the same way.

\begin{figure}[h]
	\includegraphics[height=0.25\textwidth]{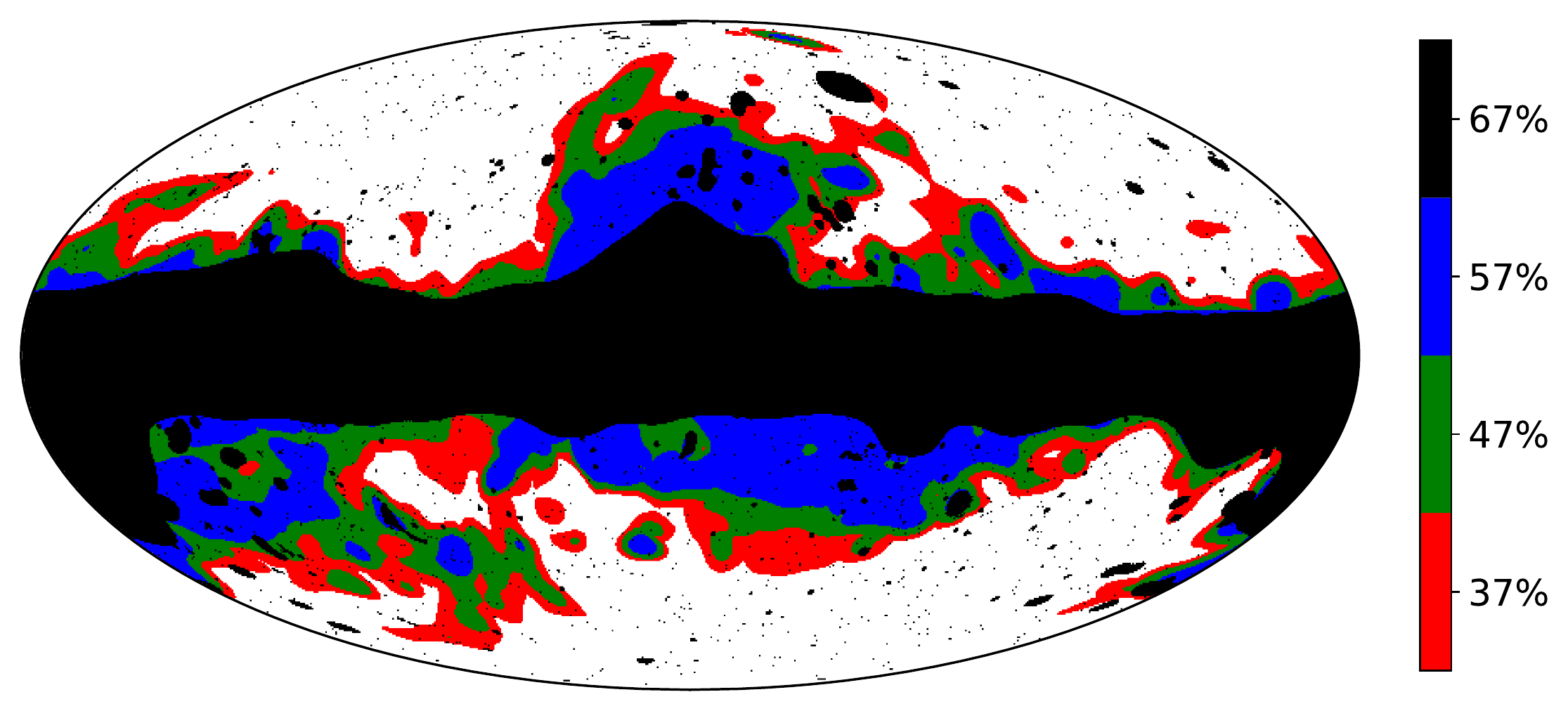}
	\caption{Masks used to test foreground contamination. Black corresponds to the fiducial mask used in the \planck\ lensing analysis, coloured masks layered below are generated from a threshold cut in the smoothed intensity of emission in 353 GHz. By construction, the larger masks are a superset of the smaller ones, so that $\mathrm{M37} \subseteq \mathrm{M47} \subseteq \mathrm{M57} \subseteq \mathrm{M67}$.}\label{fig:fsky_353}
\end{figure}

We now consider the foreground cleaned \smica\ maps, for which we compare \dxtw and \npipe data. It is helpful to consider both of these data (and corresponding simulation) products since it enables us to identify possible systematic effects arising from the map-making procedure that could lead to a false detection.
When comparing these data we perform modulation reconstruction for a variety of different sky fractions, the masks of which are shown in \figref{fig:fsky_353}. By analysing the response of the modulation reconstruction to lesser sky coverage, we check for possible galactic foreground contaminants. The baseline mask maintains $f_\mathrm{sky}\sim67\%$, and is almost identical to that used in the \planck\ lensing constraints. Additional masks are then obtained by taking a threshold cut in the 353 GHz intensity emission (after smoothing the data to avoid removing regions of high-intensity CMB). We vary the threshold in order to obtain masks with an incremental 10\% reduction in $f_\mathrm{sky}$.

For \npipe analyses we mask a small number of additional pixels, which correspond to the pre-processing mask, applied during our implementation of the \smica\ map making process; this essentially leaves the sky fraction unchanged. For \dxtw analyses (when we use the \texttt{HM} splits) the data contains missing pixels, therefore we apply further masking which results in a $1-3\%$ reduction in $f_\mathrm{sky}$.
In the following text, these masks will be referred to as M67, M57, M47 and M37, where any additional masking should be interpreted from the data it is applied to and the number convention indicates the sky fraction of the baseline masks in \figref{fig:fsky_353}.

\begin{figure}[h]
	\includegraphics[width=0.99\textwidth]{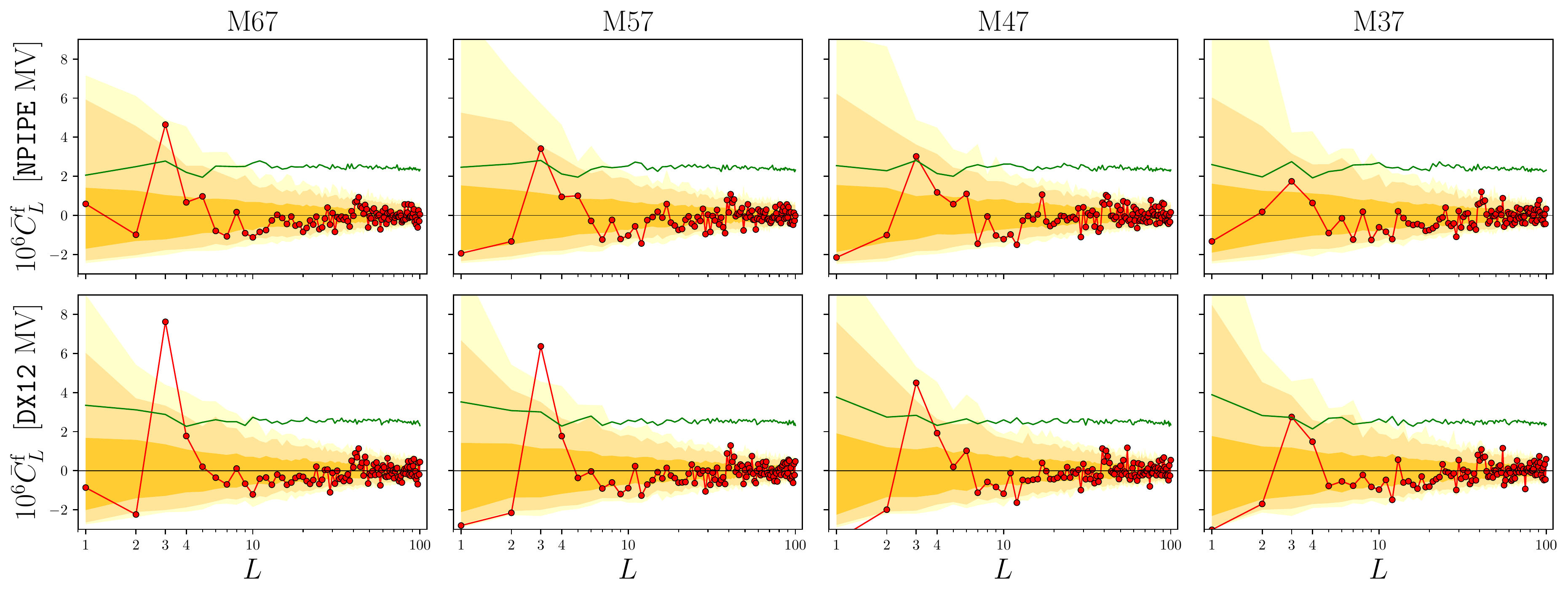}
	\caption{Uncalibrated modulation power spectra from \npipe (\texttt{A}, \texttt{B}) and \dxtw (\texttt{HM1}, \texttt{HM2}) joint-reconstructions. In each panel we plot the MV reconstruction, and from left-to-right reduce the sky fraction; corresponding to M67, M57, M47 and M37 as shown in \figref{fig:fsky_353}. The 68\%, 95\% and 99\% confidence intervals are represented by the orange contours.
}\label{fig:modf_cls_fsky_masks}
\end{figure}

Although \npipe has more end-to-end simulations available than \ffpten, to keep these initial comparisons on equal footing we construct mean field and $\rdnzerobar$ corrections using the same number of simulations (80 for mean field, 220 for $\rdnzerobar$).
In \figref{fig:modf_cls_fsky_masks} we present the results for \clmodfbar\ computed from minimum variance modulation reconstruction. In the top and bottom rows we plot the power spectra of joint reconstructions \npipe (\texttt{A}, \texttt{B}) and \dxtw (\texttt{HM1}, \texttt{HM2}) respectively, and left-to-right the reduced $\fsky$. For a given baseline mask, \npipe and \dxtw have similar spectra, with the amplitude of the large-scale reconstruction modes of \npipe being generically lower.
All of the spectra have a shape consistent with the (143, 217) GHz reconstruction, in which most of the power is found in the octopole; which has decreasing power as a function of $\fsky$.
This systematic decrease in the octopole power suggests that there may be a galactic foreground contaminant in the modulation signal, which is partially removed as the size of the galactic mask is increased. However, as we decrease the available sky fraction the statistical significance of the reconstructed signal decreases; as can be seen by the error bars expanding left-to-right.
We do not present the TT reconstruction spectra since these are essentially the same as the MV, though the spectra and errors have a slightly larger amplitude.

In the upper part of Table~\ref{tab:ivw_estimates} we provide estimates for the uncalibrated modulation power spectrum reconstructions. For each power spectrum we compute an uncalibrated $\bar{\tau}_\mathrm{NL}$ estimate via direct application of Eq.~\eqref{eq:finite_estimator}, and provide a rough estimate of the error $\sigma(\bar{\tau}_\mathrm{NL})$ from applying the same estimator to reconstructed modulation spectra of without-signal simulations and computing the resulting standard deviation. For each spectrum we form an estimate with $L_\mathrm{min}=1$ and consider $L_\mathrm{max}=2, 10$ and $50$. We provide results for both TT and MV reconstructions.

\input{sections/results/table_inv_var_weighted_estimates.tex}

When only $L\in[1, 2]$ define the estimator we sometimes have large negative estimates, which are driven directly by the negative mode(s) in the reconstructed power spectra;
these estimates are un-physical. For the remaining estimators when $L_\mathrm{max}\in[10, 50]$ we see that all of the data estimates are consistent with $\taunl=0$.
It should be noted that using the standard deviation of $\bar{\tau}_\mathrm{NL}$ from simulations is not the most rigorous approach to estimating the error, since the distribution is slightly skewed by the $\chi^2$ properties of the reconstruction modes. Our estimated error is however still useful as an approximate means of assessing the data consistency with $\taunl=0$.

The error bars for the $\bar{\tau}_\mathrm{NL}$ constraints in Table~\ref{tab:ivw_estimates} nearly saturate by $L_{\rm max}=10$, since the bulk of the signal is in the lowest multipoles. However, the measured values mostly shift downward moving from $L_{\rm max}=10$ to $L_{\rm max}=50$, by an amount that look anomalous given the little information that should be available in that range (hence the error bars being almost unchanged). This shift is dominated by the multipole range  $10 \lesssim L \lesssim 40$, corresponding to the run of mostly negative modulation power multipole estimates clearly seen by eye in e.g. Fig.~\ref{fig:modf_cls_fsky_masks}.
However, we find consistency with the simulations to $\lesssim 3\sigma$ by applying Eq.~\eqref{eq:finite_estimator} to each of the sub-intervals $5 + \delta L \leq L \leq 15 + \delta L$ for $\delta L \in [0, 10, 20, 30]$. We have also found that \npipe (\texttt{A}, \texttt{B}) is just over $3\sigma$ low for $9 \leq L \leq 35$.
While visually striking, negative signals cannot be produced by a physical modulation signal, and this multipole range is chosen a posteriori, so we do not regard this is significant evidence of a problem.

In addition to the data provided in the table, we report
$\bar{\tau}_\mathrm{NL} \pm \sigma(\bar{\tau}_\mathrm{NL}) =$
$-150 \pm 579$, $173 \pm 577$ and $156 \pm 578$
for $L_\mathrm{min}=1$ with $L_\mathrm{max} = 2, 10$ and $50$,
from the TTxPP \texttt{NPIPE} (\texttt{A}, \texttt{B}) reconstruction spectrum using the M67 mask. As discussed in Section~\ref{sec:data_and_sims}, estimators of this type have a large variance, and this can be seen directly by comparing the error on this result to the analogous case of MV \texttt{NPIPE} (A, B) given in Table~\ref{tab:ivw_estimates}, which is more than $2\times$ smaller.

\begin{figure}[h]
	\includegraphics[width=0.99\textwidth]{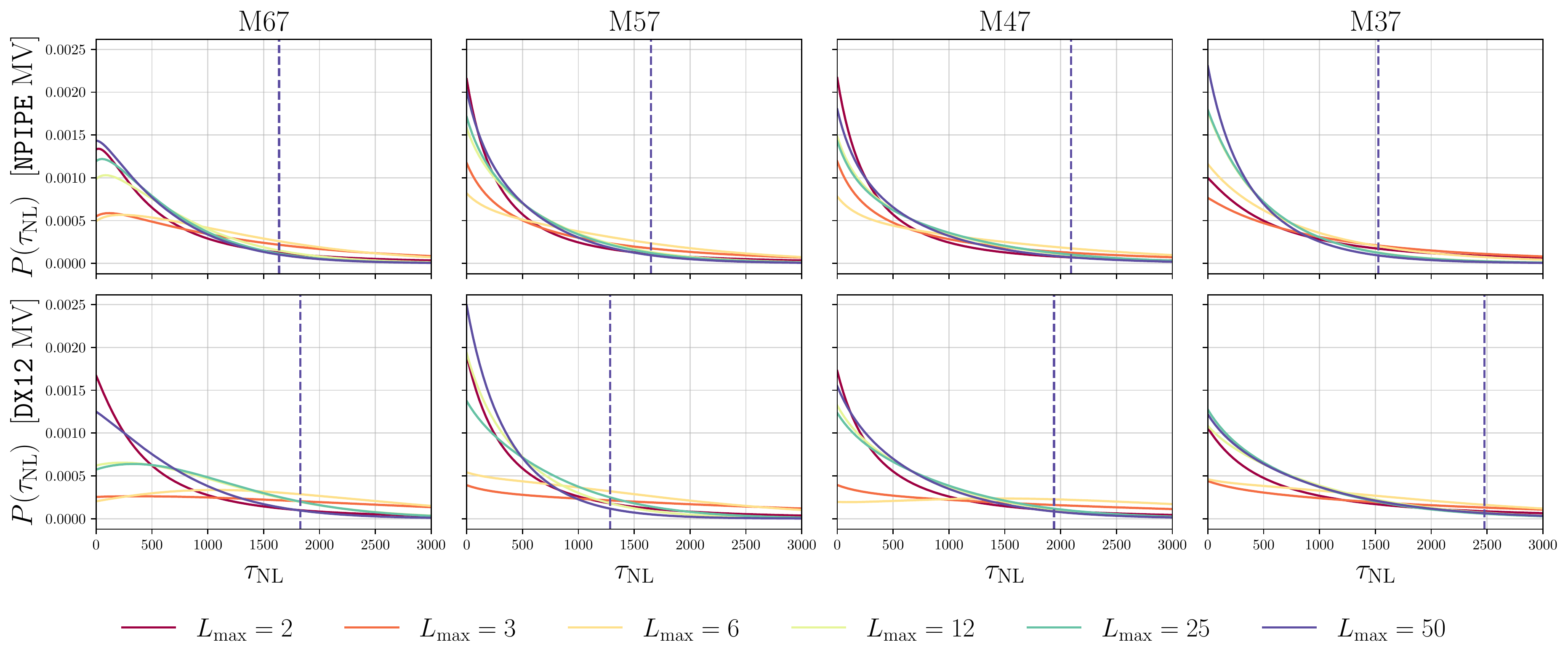}
	\caption{Uncalibrated posterior distribution of $\taunl$ estimates, from MV reconstructed modulation power spectra from \npipe (\texttt{A}, \texttt{B}) and \dxtw (\texttt{HM1}, \texttt{HM2}) data. The underlying data power spectra correspond to those given in \figref{fig:modf_cls_fsky_masks}. All posterior distributions are computed from $L=1$ to an $L_\mathrm{max}$, as indicated by the colour of the line. The single dashed line in each panel indicates the corresponding 95\% C.L. $\taunl$ constraint obtained by integrating the $L_\mathrm{max}=50$ posterior distribution.
}.\label{fig:posterior_summary}
\end{figure}

In \figref{fig:posterior_summary} we show the corresponding posterior distribution of estimates for a range of $L_\mathrm{max}$ with fixed $L_\mathrm{min}=1$. All of the posterior distributions are consistent with $\taunl=0$ with very long tails. In a few instances, the posterior peaks slightly away from zero, but in most cases peaks at zero.
The dashed vertical lines in each panel indicate the upper bound on $\taunl$ inferred from the 95\% C.L. on the $L_\mathrm{max}=50$ posterior distribution. This bound shifts as a function of $\fsky$ but is relatively stable for the larger-sky fractions. Note that these posteriors have no calibration factors of $k_L$, but do contain the Monte Carlo and Hartlap corrections to their covariances as described in Section~\ref{sec:taunl_likelihood}. Without applying the full calibration these estimates are slightly too small.

All of the uncalibrated posterior constraints are provided in the upper portion of Table~\ref{tab:posterior_constraints}. We provide the constraints from using data up to $L_\mathrm{max}=2, 10, 50$ in both temperature and polarization (except for \dxtw (143, 217) GHz). In general, the upper-bound decreases as more multipoles are included in the likelihood analysis; with the tightest constraints arising when we preserve the larger sky-fractions. This latter trend is associated with the increased (co-)variance of modulation estimators as the sky fraction is decreased, leading to broader posteriors. For M67 and M57, we find that the 95\% upper bound lies approximately between 1500-2000.

\input{sections/results/table_posterior_constraints.tex}

We finally produce results for the calibrated analyses of TT and MV modulation reconstruction using \dxtw (\texttt{HM1}, \texttt{HM2}) and \npipe (\texttt{A}, \texttt{B}) map pairs. For the calibrated \npipe analysis, we have now used all available simulations; 120 and 480 simulations for the mean field and noise bias estimates respectively. In the lower part of Table~\ref{tab:ivw_estimates} we provide the inverse-variance weighted estimates for these data. For the $L_\mathrm{max}=2$ constraint, we see that the estimate is pulled more negative compared to the uncalibrated result. This is due to the dipole and quadrupole receiving multiplicative corrections with $k_L>1$ (see inset on \figref{fig:calibrated_cls}) and the fact that the reconstructed spectra are negative for these two modes. Once further multipoles are included, the estimates become more consistent with zero. For all of the calibrated analyses the error estimate becomes slightly larger; which reflects the high weighting that the low multipoles receive for this particular form of the estimator and, as previously noted, that $k_L>1$ on these scales.

\begin{figure}
	\includegraphics[width=0.99\textwidth]{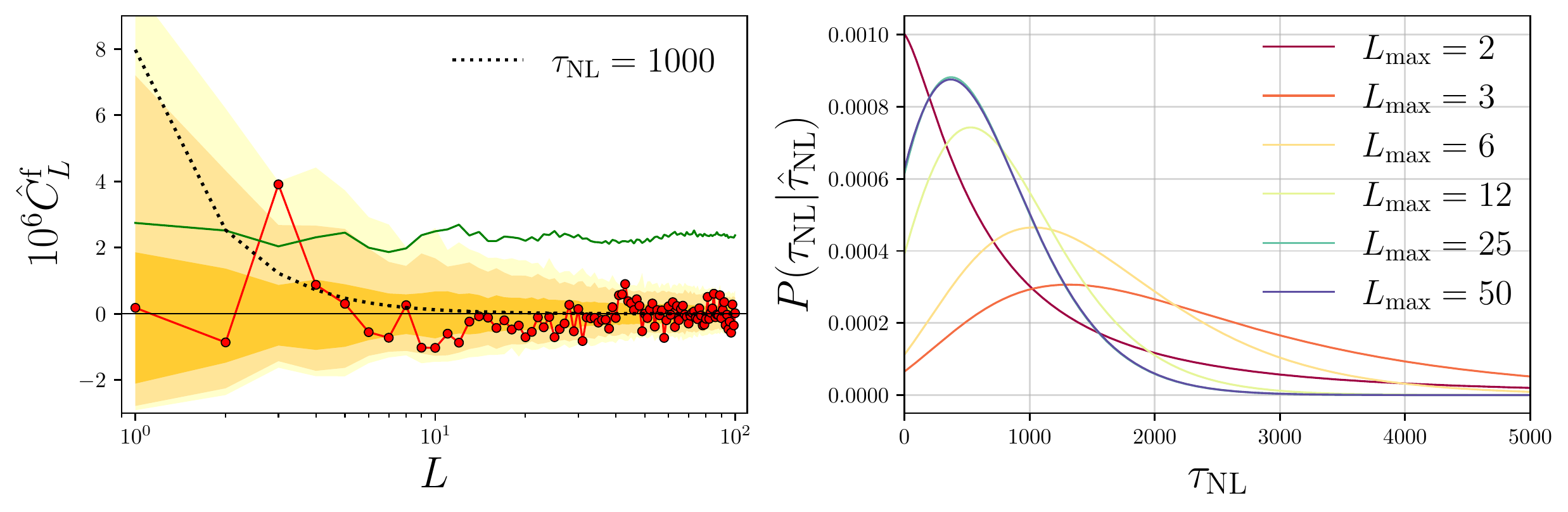}
	\caption{\textit{Left:} Calibrated MV modulation power spectrum reconstruction for \npipe (\texttt{A}, \texttt{B}) analysis. \textit{Right:} Corresponding posterior distribution of $\taunl$ estimates. This figure represents the key data for our baseline constraints.
 }\label{fig:baseline}
\end{figure}

In Table~\ref{tab:posterior_constraints} we present the calibrated $\taunl$ constraints on the posterior probability, which exhibit relatively large changes from their respective uncalibrated constraints. This is especially notable from for \dxtw 95\% C.L. constraints for $L_\mathrm{max}=2$. For these calibrated results, the posterior peaks at zero, but has a long and shallow tail, which leads to the relatively weak constraining power. As in the uncalibrated analyses, all of the posterior distributions remain consistent with zero. After calibrating, however, the posterior peaks are pulled slightly positive, such that the normalized distributions exhibit greater shifts as a function of $L_\mathrm{max}$.
For our baseline result we advocate the calibrated MV \npipe (\texttt{A}, \texttt{B}) analysis, for which we report $\taunl < 1700$ 95\% C.L. The full-calibrated result for the modulation power spectrum and the corresponding posterior distributions are shown in \figref{fig:baseline}.

\begin{figure}[h]
	\includegraphics[width=0.65\textwidth]{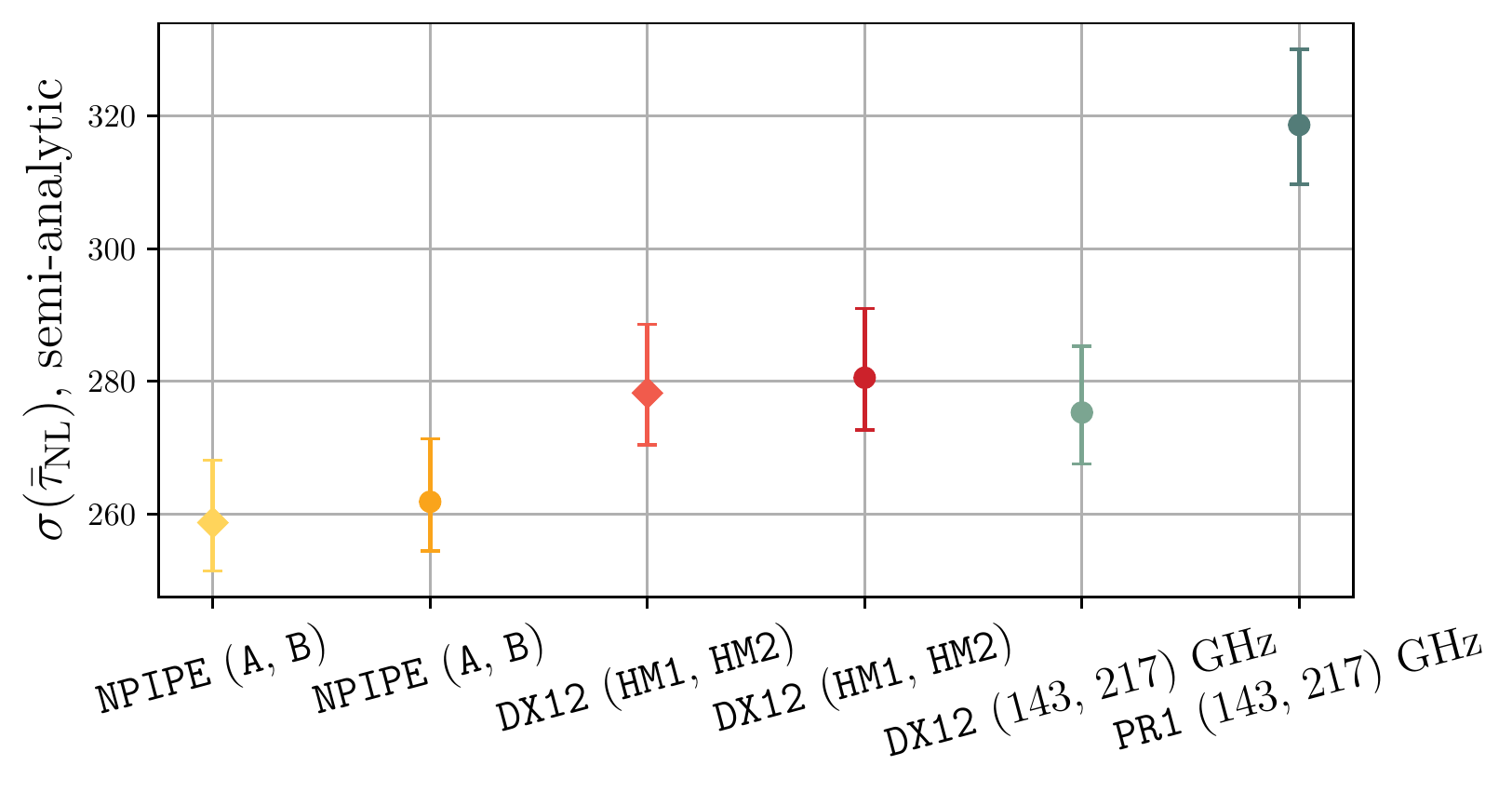}
	\caption{Semi-analytic error forecasts for $\bar{\tau}_\mathrm{NL}$ estimates with different data choices: Diamond markers correspond to MV and the remaining circular markers to TT estimators.}
	\label{fig:analytic_errs}
\end{figure}

Finally, we attempt to gauge the impact improved data quality has on $\sigma(\taunl)$ estimates using a semi-analytic approximation of the reconstruction noise bias $\nzero$. These $\nzero$ can be computed cheaply from the power spectra of the IVF maps of different data, and we do so for the \npipe, \dxtw and \Planck\ 2013 (PR1) data.
Using these spectra we simulate $10^6$ realizations of the Gaussian noise bias following a $\chi^2$ distribution, and propagate the spectra into Eq.~\eqref{eq:finite_estimator} to build $\bar{\tau}_\mathrm{NL}$ estimates.
The standard deviation from these ensembles is then used to build approximate errors for each dataset, and by further splitting the estimates into ($10^3$) subsets before computing $\sigma(\taunl)$ we estimate the sample-variance from the scatter on the errors.
The results from this exercise are given in \figref{fig:analytic_errs} for $L_\mathrm{min}=1$ and $L_\mathrm{max}=10$,
and we see that these are highly consistent with those computed in Table~\ref{tab:ivw_estimates} for the full end-to-end simulations.

Moving from PR1 to \dxtw for (143, 217) GHz analyses, there is a large improvement from the overall reduction in noise.
However when moving from TT to MV (relevant for \npipe (\texttt{A}, \texttt{B}) and \dxtw (\texttt{HM1}, \texttt{HM2})) there is a relatively small improvement, which highlights that at \planck\ polarization noise levels the reconstruction signal is largely accounted for by the temperature modes.
The forecast suggests that there is $\sim 20\%$ reduction on the error (assuming a null hypothesis) when using \npipe rather than PR1 data.
Interestingly the errors for \dxtw (143, 217) GHz seem to outperform those from \dxtw (\texttt{HM1}, \texttt{HM2}). In the analytic model, this is explained by the competing factors of noise levels and sky fraction: Although the \texttt{HM} maps have slightly lower noise properties, they require additional masking due to missing pixel data, and this few percent difference in $\fsky$ propagates directly into the error estimate.
It should further be noted that the component-separated maps have a better control over galactic and astrophysical contaminants, the details of which are not folded into this analytic calculation. The component-separated maps are therefore expected to produce cleaner estimates of primordial modulation despite the seemingly competitive reconstruction noise bias of the separate frequency maps.

%% file: sections/results/table_inv_var_weighted_estimates.tex

\begin{table}
    \centering
    \begin{tabular}{| l r | r r r | r r r | }
        \multicolumn{8}{l}{{\bf Uncalibrated inverse-variance weighted estimates:} $\bar{\tau}_\mathrm{NL} \pm  \sigma(\bar{\tau}_\mathrm{NL})$ for $1 \leq L \leq  L_\mathrm{max}$} \\

        \cline{3-8}
        \multicolumn{2}{c|}{} &  \multicolumn{3}{c|}{Temperature} &  \multicolumn{3}{c|}{Minimum variance} \\

        \cline{1-8}
        Data & Mask
        & $L_\mathrm{max}=2$
        & $L_\mathrm{max}=10$
        & $L_\mathrm{max}=50$
        & $L_\mathrm{max}=2$
        & $L_\mathrm{max}=10$
        & $L_\mathrm{max}=50$ \\

        \hline
        \hline

        \texttt{DX12} (143, 217) GHz & M67 & $-340 \pm 267$ & $-56 \pm 272$ & $-94 \pm 274$ & N/A & N/A & N/A \\

        \texttt{DX12} (\texttt{HM1}, \texttt{HM2}) & M67 & $-260 \pm 273$ & $95 \pm 274$ & $50 \pm 277$ & $-230 \pm 269$ & $113 \pm 267$ & $49 \pm 269$\\ 

        \texttt{DX12} (\texttt{HM1}, \texttt{HM2}) & M57 & $-459 \pm 301$ & $-154 \pm 302$ & $-185 \pm 306$ & $-430 \pm 302$ & $-145 \pm 294$ & $-200 \pm 297$\\ 

        \texttt{DX12} (\texttt{HM1}, \texttt{HM2}) & M47 & $-492 \pm 319$ & $-177 \pm 330$ & $-208 \pm 332$ & $-500 \pm 323$ & $-241 \pm 322$ & $-291 \pm 323$\\ 

        \texttt{DX12} (\texttt{HM1}, \texttt{HM2}) & M37 & $-384 \pm 346$ & $-166 \pm 354$ & $-207 \pm 357$ & $-425 \pm 357$ & $-302 \pm 357$ & $-369 \pm 360$\\ 

        \texttt{NPIPE} (\texttt{A}, \texttt{B}) & M67 & $-46 \pm 260$ & $129 \pm 259$ & $79 \pm 261$ & $1 \pm 248$ & $167 \pm 248$ & $107 \pm 248$\\ 

        \texttt{NPIPE} (\texttt{A}, \texttt{B}) & M57 & $-306 \pm 276$ & $-145 \pm 274$ & $-185 \pm 275$ & $-288 \pm 267$ & $-134 \pm 266$ & $-186 \pm 266$\\ 

        \texttt{NPIPE} (\texttt{A}, \texttt{B}) & M47 & $-313 \pm 282$ & $-107 \pm 286$ & $-134 \pm 287$ & $-290 \pm 273$ & $-125 \pm 278$ & $-165 \pm 279$\\ 

        \texttt{NPIPE} (\texttt{A}, \texttt{B}) & M37 & $-170 \pm 296$ & $-66 \pm 299$ & $-120 \pm 301$ & $-130 \pm 285$ & $-103 \pm 285$ & $-174 \pm 285$\\
        \hline

        \multicolumn{8}{c}{} \\

        \multicolumn{8}{l}{{\bf Calibrated inverse-variance weighted estimates:} $\hat{\tau}_\mathrm{NL}, \  \sigma(\hat{\tau}_\mathrm{NL})$ for $1 \leq L \leq  L_\mathrm{max}$} \\

        \cline{3-8}
        \multicolumn{2}{c|}{} &  \multicolumn{3}{c|}{Temperature} &  \multicolumn{3}{c|}{Minimum variance} \\

        \cline{1-8}
        Data & Mask
        & $L_\mathrm{max}=2$
        & $L_\mathrm{max}=10$
        & $L_\mathrm{max}=50$
        & $L_\mathrm{max}=2$
        & $L_\mathrm{max}=10$
        & $L_\mathrm{max}=50$ \\

        \hline
        \hline

        \texttt{DX12 HM} (\texttt{HM1}, \texttt{HM2}) & M67 & $-321 \pm 326$ & $12 \pm 317$ & $-30 \pm 319$ & $-287 \pm 321$ & $39 \pm 309$ & $-21 \pm 310$\\

        \texttt{NPIPE AB} (\texttt{A}, \texttt{B}) & M67 & $-62 \pm 290$ & $86 \pm 279$ & $36 \pm 280$ & $-63 \pm 293$ & $86 \pm 283$ & $36 \pm 283$\\

        \hline

    \end{tabular}
    \caption{Summary of the inverse-variance weighted $\taunl$ estimates. For each combination of data, mask and modulation estimator, we compute $\bar{\tau}_\mathrm{NL}$ estimates from $L=1$ to $L_\mathrm{max}$ using Eq.~\eqref{eq:finite_estimator}. Each entry in the table(s) corresponds to the data estimate and standard deviation of equivalent estimates from without-signal simulations. 
    \textit{Upper:} Uncalibrated estimates from \dxtw and \npipe data. \textit{Lower:} Calibrated estimates.}
    \label{tab:ivw_estimates}
\end{table} 

%% file: sections/results/table_posterior_constraints.tex
\begin{table}
    \centering
    \begin{tabular}{| l r | r r r | r r r | }
        \multicolumn{8}{l}{{\bf Uncalibrated $\tau_\mathrm{NL}$ constraints:} 95\% C.L. on  $P(\tau_\mathrm{NL} | \bar{\tau}_\mathrm{NL})$ for $1 \leq L \leq  L_\mathrm{max}$} \\

        \cline{3-8}
        \multicolumn{2}{c|}{} &  \multicolumn{3}{c|}{Temperature} &  \multicolumn{3}{c|}{Minimum variance} \\

        \cline{1-8}
        Data & Mask
        & $L_\mathrm{max}=2$
        & $L_\mathrm{max}=10$
        & $L_\mathrm{max}=50$
        & $L_\mathrm{max}=2$
        & $L_\mathrm{max}=10$
        & $L_\mathrm{max}=50$ \\

        \hline
        \hline

        \texttt{DX12} (143, 217) GHz & M67 & 3300 & 2600 & 2000 & N/A & N/A & N/A \\

        \texttt{DX12} (\texttt{HM1}, \texttt{HM2}) & M67 & 3400 & 3100 & 2200 & 3200 & 2600 & 1800 \\

        \texttt{DX12} (\texttt{HM1}, \texttt{HM2}) & M57 & 4100 & 2300 & 1800 & 3500 & 1800 & 1300 \\

        \texttt{DX12} (\texttt{HM1}, \texttt{HM2}) & M47 & 5200 & 4100 & 3100 & 5000 & 3000 & 1900 \\

        \texttt{DX12} (\texttt{HM1}, \texttt{HM2})  & M37 & 8700 & 3900 & 3600 & 9800 & 3000 & 2500 \\

        \texttt{NPIPE} (\texttt{A}, \texttt{B}) & M67 & 3600 & 2200 & 2000 & 3200 & 2300 & 1600 \\

        \texttt{NPIPE} (\texttt{A}, \texttt{B}) & M57 & 4000 & 2200 & 1600 & 3400 & 2200 & 1600 \\

        \texttt{NPIPE} (\texttt{A}, \texttt{B}) & M47 & 3500 & 3300 & 2600 & 3400 & 2700 & 2100 \\

        \texttt{NPIPE} (\texttt{A}, \texttt{B}) & M37 & 6200 & 2600 & 2200 & 8600 & 2100 & 1500 \\

        \hline

        \multicolumn{8}{c}{} \\

        \multicolumn{8}{l}{{\bf Calibrated $\tau_\mathrm{NL}$ constraints:} 95\% C.L. on  $P(\tau_\mathrm{NL} | \hat{\tau}_\mathrm{NL})$ for $1 \leq L \leq  L_\mathrm{max}$} \\

        \cline{3-8}
        \multicolumn{2}{c|}{} &  \multicolumn{3}{c|}{Temperature} &  \multicolumn{3}{c|}{Minimum variance} \\

        \cline{1-8}
        Data & Mask
        & $L_\mathrm{max}=2$
        & $L_\mathrm{max}=10$
        & $L_\mathrm{max}=50$
        & $L_\mathrm{max}=2$
        & $L_\mathrm{max}=10$
        & $L_\mathrm{max}=50$ \\

        \hline
        \hline
        \texttt{DX12} (\texttt{HM1}, \texttt{HM2})  & M67 & 16000 & 3900 & 2800 & 14000 & 3300 & 2300 \\

        \texttt{NPIPE} (\texttt{A}, \texttt{B})  & M67 & 5900 & 2200 & 1900 & 5800 & 2200 & 1700 \\

        \hline

    \end{tabular}
    \caption{Summary of upper limits on $\taunl$ from posterior probabilities. \textit{Upper:} Uncalibrated data constraints from \dxtw and \npipe data; \texttt{Sep.} refers to the separate frequency constraint from the (143, 217) GHz cross-estimator. \textit{Lower:} 95\% C.L. $\taunl$ constraints from calibrated analyses. The \npipe (\texttt{A}, \texttt{B}) constraint in the lower table is our baseline result.
}
    \label{tab:posterior_constraints}
\end{table} 

%% file: sections/results/conclusions.tex
\section{Conclusions}

Using the latest \planck\ data, we found the significantly tightened constraint $\taunl < 1700$ 95\% C.L., reducing the upper limit by a factor of $\sim 1.7 $ compared to previous results. A Universe with purely Gaussian primordial fluctuations remains consistent with the data, though the upper limit remains substantially larger than would be predicted in most multi-field inflation models.

In addition to making use of new data scans and polarization, and using the improved \npipe\ processing, we also made several significant improvements to the analysis methodology, including a more optimal and robust model for the connected biases and use of foreground-separated maps. The quadratic estimator pipeline is very similar to that for CMB lensing, but with the additional requirement to consistently model the expected lensing four-point signal. Several lessons from this work have already been directly useful to inform the analogous updated \npipe\ CMB lensing analysis (in prep.). The constraint is still dominated by information from the temperature due to the relatively low signal to noise in the \planck\ polarization, but could be significantly improved in future with small-scale low-noise polarization data. More dramatic improvements may be possible in conjunction with low shot noise large-scale structure surveys and small-scale kinetic SZ observations~\cite{Kumar:2022yus}.

The current analysis also could be further optimized by more optimal filtering and handling of the noise anisotropy (e.g. see Ref.~\cite{Mirmelstein:2019sxi} in the context of lensing), but given the signal is temperature dominated, the constraint is not expected to improve dramatically. Future work could also easily extend the analysis to scale-dependent \taunl-like signals.

%% file: sections/acknowledgements/acknowledgements.tex
\section*{Acknowledgements}
KM acknowledges support from STFC training grant ST/S505766/1. AL and JC acknowledge support from the European Research Council under the European Union's Seventh Framework Programme (FP/2007-2013) / ERC Grant Agreement No. [616170], and AL support by the UK STFC grants ST/P000525/1 and  ST/T000473/1. JC acknowledges support from a SNSF Eccellenza Professorial Fellowship (No. 186879). 
This research used resources of the National Energy Research Scientific Computing Center (NERSC), a U.S. Department of Energy Office of Science User Facility located at Lawrence Berkeley National Laboratory. 
KM is grateful to Reijo Keskitalo for the provision of - and information pertaining to - \planck\ simulation products available on NERSC. KM further thanks Maude Le Jeune and Jean-Francois Cardoso for helpful discussion regarding \smica\ processed simulations. 
Some of the key results from this paper were derived using the \href{https://github.com/carronj/plancklens}{\texttt{plancklens}} and \texttt{healpy} (\texttt{healpix}) \cite{Zonca2019,2005ApJ...622..759G} software packages.

%% file: sections/appendix/appendix.tex
\appendix
\input{sections/appendix/spectral_corrections}

%% file: sections/appendix/spectral_corrections.tex
\section{Spectral corrections}\label{sec:spec_correct}

\begin{figure}[h]
	\includegraphics[width=0.95\textwidth]{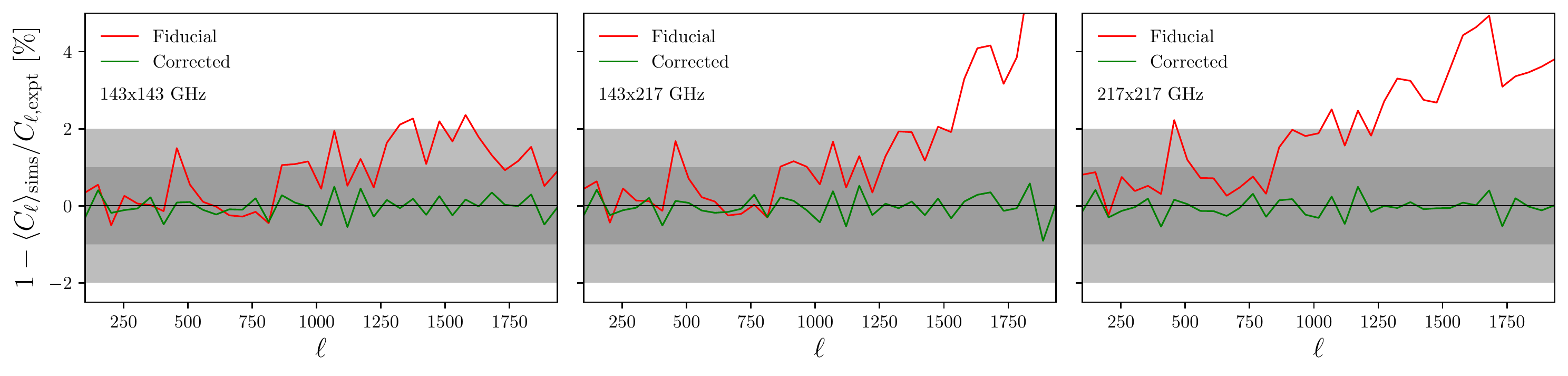}
	\caption{Temperature spectral improvements $\Delta C_\ell^{TT}$, as measured from the \dxtw 143/217 GHz data and simulations. The power spectra are computed directly from the WF maps, and have been binned in multipole intervals of $\Delta \ell = 50$ and weighted by a $L(L+1)$ kernel. In each panel the fractional difference of the fiducial and corrected maps is given in red and green respectively. For reference, two grey shaded bands indicate the 1\% and 2\% difference contours. \textit{Left:} Auto-power spectra from 143 GHz . \textit{Centre:} Cross-power spectra from 143 GHz and 217 GHz. \textit{Right:} Auto-power spectra from 217 GHz.}
	\label{fig:dcl_npipe_tt}
\end{figure}

Accurate computation of the \rdnzero spectral correction described in Section~\ref{section:mf_rdn0} rely on the Gaussian power between the data and simulations being perturbatively close. In order to ensure this, we add a small amount of Gaussian power to the simulations and/or data. We quantify this difference by measuring the power spectra of the WF data and the ensemble averaged power spectra of WF simulation maps; $\Delta C_\ell = C_{\ell, \mathrm{expt}} - \langle C_\ell\rangle_\mathrm{sims} \approx 0$. The WF maps (rather than the raw maps) provide a better benchmark for assessing the power differences since it is the filtered maps that enter the quadratic estimator. 
Moreover, in the case of e.g. the (143, 217) GHz analysis where we perform a template projection of the 857 GHz intensity map during the filtering stage, there is a reduction in power that would not otherwise be measured on the raw (unfiltered) maps.

Since we consider the joint modulation reconstruction between map pairs (e.g. see discussion towards the bottom of Section~\ref{sec:modulation_estimator}) it is important that we not only correct the Gaussian auto power spectrum within a given map set, but also the cross power between different map sets entering the quadratic estimator. By map sets, we are collectively referring to a given data and its corresponding set of simulations.
A concrete example is ensuring that $\Delta C_\ell \approx 0$ for 143x143 GHz, 217x217 GHz and 143x217 GHz simultaneously.
We achieve this by solving the system of equations
\begin{align}\label{eq:gen_deltacl}
	C_\ell[P_{\ell m}^\mathrm{a, dat}, P_{\ell m}^\mathrm{b, dat}]
	-
	\langle C_\ell[P_{\ell m}^\mathrm{a, sim}, P_{\ell m}^\mathrm{b, sim}] \rangle
	&=
	\Delta C_\ell^\mathrm{ab},\\
	C_\ell[P_{\ell m}^\mathrm{a, dat}, P_{\ell m}^\mathrm{a, dat}]
	-
	\langle C_\ell[P_{\ell m}^\mathrm{a, sim}, P_{\ell m}^\mathrm{a, sim}] \rangle
	&=
	\Delta C_\ell^\mathrm{aa},\\
	C_\ell[P_{\ell m}^\mathrm{b, dat}, P_{\ell m}^\mathrm{b, dat}]
	-
	\langle C_\ell[P_{\ell m}^\mathrm{b, sim}, P_{\ell m}^\mathrm{b, sim}] \rangle
	&=
	\Delta C_\ell^\mathrm{bb},
\end{align}
where $P_{\ell m}\in\{T_{\ell m}^\mathrm{WF}, E_{\ell m}^\mathrm{WF}, B_{\ell m}^\mathrm{WF}\}$ and `a' and `b' refer to distinct map sets. We want to construct solutions such that $\Delta C_\ell^{ab} = \Delta C_\ell^{aa} = \Delta C_\ell^{bb} = 0$. To this end, we assume the following form for the corrections in harmonic space
\begin{align}
	\text{Map set a: \ }
	&\begin{cases}
		P_{\ell m}^\mathrm{a, dat}
		\rightarrow
		&\hat{P}_{\ell m}^\mathrm{a, dat}
		=
		P_{\ell m}^\mathrm{a, dat}
		+
		\theta_{\ell m}^\mathrm{a}
		+
		\phi_{\ell m} \\
		P_{\ell m}^\mathrm{a, sim}
		\rightarrow
		&\hat{P}_{\ell m}^\mathrm{sim, a}
		=
		P_{\ell m}^\mathrm{a, sim}
		+
		\delta_{\ell m}^\mathrm{a}
		+
		\epsilon_{\ell m}
	\end{cases} \\
	\text{Map set b: \ }
	&\begin{cases}
		P_{\ell m}^\mathrm{b, dat}
		\rightarrow
		&\hat{P}_{\ell m}^\mathrm{b, dat}
		=
		P_{\ell m}^\mathrm{b, dat}
		+
		\theta_{\ell m}^\mathrm{b}
		+
		\phi_{\ell m} \\
		P_{\ell m}^\mathrm{b, sim}
		\rightarrow
		&\hat{P}_{\ell m}^\mathrm{sim, b}
		=
		P_{\ell m}^\mathrm{b, sim}
		+
		\delta_{\ell m}^\mathrm{b}
		+
		\epsilon_{\ell m}
	\end{cases}
\end{align}
where all of the additive corrections are only correlated with themselves. This means that (by construction) $\phi_\ell$ and $\epsilon_\ell$ fix the cross map set power spectra difference, and the remaining the auto map set power difference.

Substituting the corrected harmonics into our ansatz Eq.~\eqref{eq:gen_deltacl}, the power difference can be written in terms of the isotropic power spectrum of the harmonic corrections
\begin{align}
	\Delta C_\ell^\mathrm{ab} &= (C_\ell^\epsilon - C_\ell^\phi), \\
	\Delta C_\ell^\mathrm{aa} &= (C_\ell^\epsilon - C_\ell^\phi) + (C_\ell^{\delta, \mathrm{a}} - C_\ell^{\theta, \mathrm{a}}), \\
	\Delta C_\ell^\mathrm{bb} &= (C_\ell^\epsilon - C_\ell^\phi) + (C_\ell^{\delta, \mathrm{b}} - C_\ell^{\theta, \mathrm{b}}).
\end{align}
We can obtain solutions in terms of the combinations given in parentheses (...) by treating them as single variables. The solutions can then be unpacked further by requiring power spectra to be non-negative via a piecewise definition, e.g.
\begin{align}
	C_\ell^\phi &=
		\begin{dcases}
		| (C_\ell^\epsilon - C_\ell^\phi) | \quad &\text{if} \quad (C_\ell^\epsilon - C_\ell^\phi) < 0 \\
		\quad 0 \quad &\text{if} \quad (C_\ell^\epsilon - C_\ell^\phi) \geq 0 \\
		\end{dcases}, \\
	C_\ell^\epsilon &=
		\begin{dcases}
		\phantom|(C_\ell^\epsilon - C_\ell^\phi)\phantom| \quad &\text{if} \quad (C_\ell^\epsilon - C_\ell^\phi) \geq 0 \\
		\quad 0 \quad &\text{if} \quad (C_\ell^\epsilon - C_\ell^\phi) < 0 \\
		\end{dcases}.
\end{align}
The remaining components $C_\ell^{\delta, \mathrm{a}}$, $C_\ell^{\theta, \mathrm{a}}$, $C_\ell^{\delta, \mathrm{b}}$ and $C_\ell^{\theta, \mathrm{b}}$ are obtained in the same way. Since the solutions for the corrective power spectra depend on the empirical power difference, they are in general not smooth functions of $\ell$. We therefore bin the solutions and interpolate over $\ell$ in order to construct smooth continuous functions.
\footnote{A bin width of $\Delta \ell = 51$ is used throughout. Note that to minimize boundary effects incurred during binning and interpolation, we use the multipole range $30 \leq\ell\leq 2048$, then retain multipoles $100 \leq \ell \leq 2018$ for final analyses.}

Recall that these spectral corrections are built from the power spectra of WF maps. We want the equivalent result for unfiltered maps on the full sky, such that we can simulate and apply corrections to the raw data and simulations before reprocessing. The required transformation is (approximately) given by
\begin{equation}\label{eq:wf2unfiltered}
	C_\ell^\mathrm{Raw} \simeq C_\ell^\mathrm{WF}\left(1 + \frac{N_\ell}{C_\ell^\mathrm{Fid.}} \right)^2 \frac{\mathcal{B}_\ell^2}{\fsky},
\end{equation}
where $\mathcal{B}_\ell$ is the beam transfer function and $N_\ell=\Delta^2 \mathcal{B}_\ell^{-2}$ is the approximate isotropic noise power spectrum; for a fixed noise level $\Delta$. For uncorrelated corrections, we can simulate the corresponding isotropic map corrections directly from Eq.~\eqref{eq:wf2unfiltered} using \texttt{healpix}. For correlated components, we should instead simulate harmonic coefficients from the WF spectra, then transform these via an appropriate definition of the instrumental beam and noise (if they differ) for each map set;
\begin{equation}
	a_{\ell m}^\mathrm{Raw} \simeq a_{\ell m}^\mathrm{WF}\left(1 + \frac{N_\ell}{C_\ell^\mathrm{Fid.}} \right) \frac{\mathcal{B}_\ell}{\sqrt{f\mathrm{sky}}},
\end{equation}
before finally transforming the harmonics into real space maps. 